\documentclass[%
 reprint, prapplied, aps,
]{revtex4-2}

\usepackage{graphicx}
\usepackage{color}
\usepackage{placeins}
\usepackage{hyperref}
\usepackage{tabularx,colortbl}
\usepackage{graphicx}
\usepackage{placeins}
\usepackage{hyperref}
\usepackage{tabularx,colortbl}
\usepackage{comment}
\usepackage{amsmath,amssymb,amsfonts}
\usepackage{algorithmic}
\usepackage{array}
\usepackage{textcomp}
\usepackage{threeparttable}
\usepackage{amsmath}
\usepackage{bm}
\usepackage{float}
\usepackage{dcolumn}

\begin{document}
%

\title{Dual-Polarized All-Metallic Metagratings for Perfect Anomalous Reflection}

\author{Oshri Rabinovich}
 \email{oshrir@campus.technion.ac.il}

\author{Ariel Epstein}%
\email{epsteina@ee.technion.ac.il}

\affiliation{Andrew and Erna Viterbi Faculty of Electrical Engineering, Technion - Israel Institute of Technology, Haifa 3200003, Israel.}%


%



\begin{abstract}

We theoretically formulate and experimentally demonstrate the design of metagratings (MGs) composed of periodic rectangular grooves in a metallic medium, intended for perfect anomalous reflection. 
Using mode matching, a semianalytical scheme for analysis and synthesis of such MGs, containing multiple, arbitrarily arranged, grooves per period, is derived.
Following the typical MG design approach, we use this formalism to identify the relevant Floquet-Bloch (FB) modes and conveniently formulate constraints for suppression of spurious scattering, directly tying the structure's geometrical degrees of freedom (DOFs) to the desired functionality. Solving this set of constraints, in turn, yields a detailed fabrication-ready MG design, without any full-wave optimization.
Besides providing means to realize highly-efficient beam deflection with all-metallic formations, we show that the rectangular (two-dimensional) groove configuration enables \emph{simultaneous} manipulation of both transverse electric (TE) and transverse magnetic (TM) polarized fields, unavailable to date with common, printed-circuit-board-based, microwave MGs. In addition, we highlight a physical limitation on the TE-polarization performance, preventing the ability to achieve perfect anomalous reflection in any desired angle.
These capabilities are verified using three MG prototypes, produced with standard computer numerical control (CNC) machines, demonstrating both single- and dual-polarized control of multiple diffraction modes.
These results enable the use of MGs for a broader range of applications, where dual-polarized control is required, or 
all-metallic devices are preferable (e.g., spaceborne systems or at high operating frequencies).

\end{abstract}



\maketitle

%

\section{Introduction}

Metagartings (MGs) have been attracting much interest in the last few years \cite{sell2017large,memarian2017wide,ra2017meta,epstein2017unveiling,wong2018perfect,fan2018perfect,rabinovich2018analytical,popov2018controlling,rabinovich2019arbitrary}. 
These devices, which consist of periodic, typically-sparse, arrangements of polarizable particles (meta-atoms), allow efficient realization of a variety of beam manipulation functionalities using relatively simple structures and semianalytical design procedures. In particular, compared to common metasurfaces (MSs) \cite{glybovski2016metasurfaces,achouri2018design}, they offer a solution to the implementation difficulties associated with the dense closely-packed meta-atom distributions required to sustain homogenization therein, forming a 
promising concept for many scientific endeavours and engineering applications.

\begin{figure*}[t]
\centering
\includegraphics[width=6in]{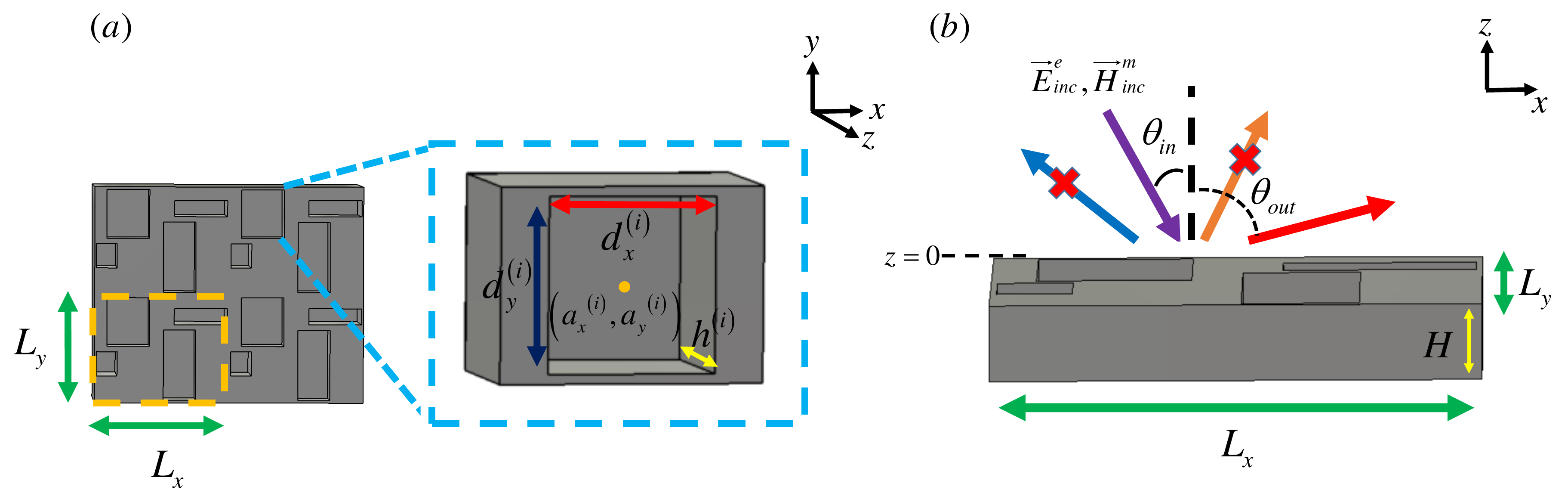}
\caption{Physical configuration of the proposed all-metallic MG. (a) Front view, showing multiple, arbitrarily arranged, grooves per period. The period is marked by an orange dashed rectangle, with periodicities $L_x$ and $L_y$ along the two lateral dimensions. Inset: the $i$th groove within the period is characterized by its center coordinates, width, height, and depth, all can serve as potential geometrical DOFs for the design procedure. (b) Side view, describing a general dual-polarized anomalous reflection scenario, where a TM ($\vec{H}^m_\mathrm{inc}\propto\hat{y}$) or TE ($\vec{E}^e_\mathrm{inc}\propto\hat{y}$)  plane wave impinges the MG with an angle of incidence $\theta_{\mathrm{in}}$ and is efficiently reflected towards a prescribed non-specular direction $\theta_\mathrm{out}$, while suppressing coupling to other propagating FB modes.}
\label{Fig:Configuration}
\end{figure*}

Recent work on MGs has covered a major portion of the electromagnetic spectrum; from microwave applications \cite{ra2017meta,molero2020planar,epstein2017unveiling,wong2018perfect,rabinovich2018analytical,popov2018controlling,rabinovich2019arbitrary}, through millimeter wave frequencies (mmWave) \cite{dong2019low} to the terahertz (THz) and optical regimes \cite{sell2017large,chalabi2017efficient,fan2018perfect,deng2018facile,behroozinia2020real}. The concept was found to be useful also in other physical fields governed by wave phenomena, e.g. acoustics 
\cite{fu2019reversal,hou2019highly,packo2019inverse} and even quantum mechanics \cite{shahmoon2017cooperative,bekenstein2020quantum}.
Within this wide range of disciplines, MGs have been shown to effectively tackle a variety of functionlaities, such as anomalous reflection \cite{ra2017meta,chalabi2017efficient,memarian2017wide,epstein2017unveiling,wong2018perfect,rabinovich2018analytical,dong2019low,rabinovich2019experimental,popov2018controlling,behroozinia2020real}, anomalous refraction \cite{sell2017large,epstein2018eucap,fan2018perfect,packo2019inverse} and focusing \cite{rabinovich2019arbitrary,kang2020efficient}. More recently, reconfigurable MGs for instantaneous control over the scattering characteristics were also introduced \cite{tian2017electronically,ra2018reconfigurable,casolaro2019dynamic}, and space-time modulated metagratings were suggested as means to realize advanced nonreciprocal responses \cite{hadad2019space,ra2020nonreciprocal}.

The meta-atom properties have a crucial effect on the MG performance in general, and on its response to polarized fields in particular. Analytically designed MGs, mainly demonstrated at microwave frequencies based on printed-circuit-board (PCB) technology, relied almost exclusively on loaded wire meta-atoms, susceptible only to transverse electric (TE) polarized fields
\cite{epstein2017unveiling,rabinovich2018analytical,popov2018controlling,popov2019constructing,rabinovich2019arbitrary}. MGs based on magnetically-polarizable particles in the form of conducting loops were envisioned in other reports \cite{ra2017meta, popov2019designing}, with potential for diffraction engineering of transverse magnetic (TM) beams.  However, such geometries require modifications to fit standard fabrication techniques and have not been demonstrated to date. 
Dual-polarized MGs, namely, with the ability to realize efficient beam deflection of both TE and TM polarized incident waves, scarcely appear in the literature, and have mainly relied on numerical optimization of the entire macro-period \cite{sell2017large,callewaert2018inverse}.
Indeed, to the best of our knowledge, a rigorous analytical methodology for designing MGs naturally enabling simultaneous control of dual-polarized fields at microwave frequencies has not been presented to date. 

Moreover, the current designed and fabricated MGs mostly rely on dielectric substrates, be it in the microwave \cite{wong2018binary,popov2019constructing,rabinovich2018analytical}, mmWave \cite{dong2019low} or optical \cite{fan2018perfect,deng2018facile} regimes. However, among the variety of applications in which MG devices can be utilized, systems intended for space exploration or communication satellites are of great importance
\cite{gonzalez2018additive,de2019manufacturing}. For such purposes, devices should sustain extreme temperature conditions and properly operate under high levels of radiation, making the utilization of relatively sensitive dielectric materials highly undesirable. 
Instead, all-metallic structures should be used. 

In this paper, we fill these gaps, considering an all-metallic dual polarized MG configuration in the form of a metal slab with periodically distributed rectangular grooves of finite volume (Fig. \ref{Fig:Configuration}). For maximum versatility, we consider the number of grooves in a period and the individual groove dimensions, namely, their width, height, and depth, as degrees of freedom (DOFs) in our design. As the grooves are actually short-circuited sections of rectangular waveguides, they can principally couple power to and from TE and TM modes (propagating and evanescent alike) \cite{balanis2012advanced}. Tuning the groove dimensions, thus, affects the scattering of both TE and TM polarized fields, paving the path to the desired dual-polarized diffraction control.
Based on this rationale, we formulate a semianalytical solution to the TE and TM scattering problem for such a configuration, combining Floquet-Bloch (FB) theory with the mode-matching technique, directly tying the groove geometry and distribution to the scattered fields. Subsequently, we follow the conceptual MG synthesis methodology \cite{ra2017meta,epstein2017unveiling,rabinovich2018analytical,rabinovich2019arbitrary} to obtain the desired functionality: for given diffraction requirements, we identify the required geometrical DOFs, and express
the corresponding constraints using the field quantities obtained from the model. 
The set of constraints is resolved using standard library functions in MATLAB, similar to \cite{rabinovich2019arbitrary}, finally yielding the detailed fabrication-ready all-metallic MG design, without resorting to full-wave simulations.

This proposed structure can be perceived as a generalization of configurations known as 
surface relief gratings, investigated in the past based on the rigorous coupled wave analysis (RCWA) technique, usually formulated for 1D groove structures (corrugations) \cite{moharam1986rigorous,gaylord1986zero}. In contrast to these previous reports, which focused on \emph{analysis} of these corrugated gratings, the work presented herein develops and demonstrates a systematic \emph{synthesis} methodology, utilizing a reliable analytical model to derive design specifications directly from high-level functionality requirements. Importantly, it considers multiple different 2D grooves in a period, and enables precise control of the coupling to 
numerous propagating FB modes. 
More recently, with the rising interest in MG-based devices, other investigations considering \emph{theoretically} all-metallic grating geometries, have been launched 
\cite{hemmatyar2019wide,rahmanzadeh2020perfect}. 
Nevertheless, the analyses presented therein are restricted to 1D grooves, manipulate only TM-polarized fields, and neglect higher-order modes within the groove-formed waveguides (only modes above cutoff are considered). In contrast, we present herein a combined theoretical and experimental effort, establishing a general synthesis procedure for dual-polarized all-metallic MGs for anomalous reflection, with an arbitrary number of (2D) rectangular grooves per period and multiple propagating FB modes: from a rigorous theoretical formulation that considers the \emph{complete} set of eigenmodes inside and outside the grooves to a comprehensive \emph{experimental} validation at microwave frequencies.
 
Indeed, to verify the theoretical derivation and demonstrate the versatility of this all-metallic MG configuration we utilize the developed methodology to design, fabricate, and experimentally characterize three prototype devices. The first device validates the MG capability of implementing wide-angle anomalous reflection for TM-polarized fields using a single groove in the period. The second experiment demonstrates that the proposed structure can handle more complex diffraction engineering tasks by utilizing multiple grooves per period, enabling highly-efficient coupling of the incident power into a single radiation channel in the presence of multiple propagating FB modes. The last MG we examine is a dual-polarized anomalous reflector, showing that a single groove in the period is sufficient to realize simultaneous deflection of TE and TM incident waves. In addition, we shed light on the performance of TE-polarized anomalous reflection achievable with the proposed device, in consistency with observations made previously with respect to MSs performing similar functionalities.
Semianalytical predictions, full-wave simulations, and experimental results agree very well, providing unambiguous validation of the concept and its practical viability.

Although these devices are expected to be somewhat thicker and heavier than their PCB counterparts, they feature two distinctive advantages: they exhibit negligible losses, and there is no load geometry that is tricky to model accurately \cite{popov2019designing}.
Overall, the proposed design scheme facilitates incorporation of MGs in applications requiring simultaneous control of dual-polarized fields \cite{encinar2004three,encinar2006dual,zheng2014highly,khorasaninejad2015efficient}, especially where all-metallic constructs are preferable \cite{gonzalez2017design,gonzalez2018ka,alex2020wave}. 


\section{Theory}
\label{sec:theory}
\subsection{Formulation}
\label{subsec:formulation}
We consider a periodic configuration of rectangular grooves made in a metallic medium [modelled as a perfect electric conductor (PEC)] filling the half-space $z<0$ (Fig. \ref{Fig:Configuration}) \footnote{In practice, an overall slab thickness of $H$ that is larger than the maximal groove depth $h^{(i)}$ by at least several skin depths will be sufficient to retain the model validity for realistic implementations.}. The periodicities along the $x$ and $y$ axes, respectively, are $L_{x}$ and $L_{y}$. 
Each such period contains $N_{\mathrm{grv}}$ grooves, the dimensions and position of which can be set at will, in principle. 
Correspondingly, we denote the center of the $i$th groove as $(a^{(i)}_{x},a^{(i)}_{y})$, its width (in the $x$ direction) as $d^{(i)}_{x}$, its height (in the $y$ direction) as $d^{(i)}_{y}$, and its depth as $h^{(i)}$ [Fig. \ref{Fig:Configuration}(a)]; the boundaries of the $i$th groove are thus $x^{(i)}_{\pm}=a^{(i)}_{x}\pm d^{(i)}_{x}/2$ and $y^{(i)}_{\pm}=a^{(i)}_{y}\pm d^{(i)}_{y}/2$.
A TE ($E_z=0$) or TM ($H_z=0$) polarized plane wave impinges upon the MG configuration with an angle of incidence $\theta_\mathrm{in}$ relative to the normal, as shown in Fig. \ref{Fig:Configuration}(b). The plane of incidence is the $\widehat{xz}$ plane; however, the formalism can be readily modified to accommodate excitations from other planes of incidence as well.

The incident electric and magnetic fields can be written for the case of TE (superscript $e$) or TM (superscript $m$) excitation, respectively, as \footnote{Throughout the paper, quantities associated with the TE- (TM-) polarized excitation are denoted with $e$ ($m$) superscripts.
}
\begin{equation}
\label{eq:E_H_inc}
\begin{aligned}
\Vec{E}^{e}_{\mathrm{inc}}=\hat{y}E^{e}_{0}e^{-jk\sin\theta_{\mathrm{in}}x}e^{jk\cos\theta_{\mathrm{in}}z}\\
\Vec{H}^{m}_{\mathrm{inc}}=\hat{y}H^{m}_{0}e^{-jk\sin\theta_{\mathrm{in}}x}e^{jk\cos\theta_{\mathrm{in}}z},
\end{aligned}
\end{equation}
where $E^e_0$, $H^m_0$ are the complex amplitudes of the incoming TE or TM plane waves, $k=\omega\sqrt{\mu\varepsilon}$ and $\eta=\sqrt{\mu/\varepsilon}$, respectively, are the wavenumber and wave impedance in the surrounding medium (vacuum as a default), having permittivity $\varepsilon$ and permeability $\mu$
; the harmonic time dependency is $e^{j\omega t}$.

Our goal is to design a MG that would funnel all the incoming power from either TE or TM polarized waves (or both) into specific FB harmonics following a desired partition. To this end, we express the reflected fields using their FB expansion, the fields in the grooves using the relevant waveguide eigenmodes, and use mode matching on the aperture $z=0$ to retrieve the scattering coefficients.
Formulating constraints on the latter, manifesting the relations between the groove configurations and the coupling to the various FB channels, would eventually allow retrieval of the MG geometry (groove distribution and dimensions) to implement the design goal.

Consequently, we derive the reflected fields at $z>0$, solving the problem for each of the excitation fields $\Vec{E}_{\mathrm{inc}}^{e}$ and $\Vec{H}_{\mathrm{inc}}^{m}$ separately, in consistency with the superposition principle.
Considering the periodicity of the structure, the $z$ components of the electric and magnetic reflected fields ($z\geq 0$) \footnote{While both electric and magnetic $z$ components will appear for both polarized excitations near the aperture ($z\rightarrow0$) due to the symmetry-breaking finite-size grooves, one of them ($E_z$ in the TE case and $H_z$ in the TM case) will vanish in the far-field ($z\rightarrow\infty$), allowing proper distinction between TE- and TM-polarized scattered fields.} for each polarization can be expressed as FB mode expansions, following the FB theorem \cite{schachter1997periodic}, reading
\begin{equation}
\label{eq:E_H_ref_z}
\begin{aligned}
\!\!\!&E^{\{e,m\}}_{\mathrm{ref},z}\!\!=\!\!\!\!\!\sum_{n_x=-\infty}^{\infty}\sum_{n_y=-\infty}^{\infty}\!\!\!\!\eta A^{\{e,m\}}_{n_{x}n_{y}} e^{-jk^{(n_x)}_x x}e^{-jk^{(n_y)}_{y}y}e^{-jk_{z}z}\\
\!\!\!&H^{\{e,m\}}_{\mathrm{ref,z}}\!\!=\!\!\!\!\!\sum_{n_x=-\infty}^{\infty}\sum_{n_y=-\infty}^{\infty} \!\!\!\! B^{\{e,m\}}_{n_{x}n_{y}}e^{-jk^{(n_x)}_{x}x}e^{-jk^{(n_y)}y}e^{-jk_{z}z}
\end{aligned}
\end{equation}
where the complex coefficients $A^{e}_{n_{x}n_{y}}$ ($A^{m}_{n_{x}n_{y}}$) and $B^{e}_{n_{x}n_{y}}$ ($B^{m}_{n_{x}n_{y}}$) correspond to the $(n_x,n_y)$-order reflected mode for the TE- (TM-) polarized excitation.
The lateral wavenumbers are given in terms of the harmonic index and its projection on the first Brillouin zone, namely, $k^{(n_x)}_{x}=k\sin\theta_{\mathrm{in}}+\frac{2\pi}{L_{x}}n_{x}$ and $k^{(n_y)}_{y}=\frac{2\pi}{L_{y}}n_{y}$. For brevity, we will denote from now on the lateral wavenumbers as $k_{x}$ and $k_{y}$ for the $x$ and $y$ directions, respectively. The longitudinal wavenumber is subsequently retrieved from the dispersion relation, reading $k_{z}^{2} = k^{2}-k_{x}^{2}-k_{y}^{2}$ \footnote{In order to satisfy the radiation condition at $z\to \infty $, the imaginary part of $k_{z}$ should be non-positive when choosing the branch of the square root.}. 

Next, the fields inside each groove ($z\leq 0$) are formulated by imposing the boundary conditions at the groove PEC walls. The $z$ components of the electric and magnetic fields for the $i$th groove are thus given by \cite{balanis2012advanced}
\begin{equation}
\label{eq:TM_TE_modes_holes}
\begin{aligned}
E^{\{e,m\},(i)}_{\mathrm{grv},z}&=\sum_{m_x=1}^{\infty}\sum_{m_y=1}^{\infty}\eta^{(i)}_{d}C^{\{e,m\},(i)}_{m_xm_y}\sin\left[\frac{\pi m_x}{d^{(i)}_{x}}(x-x_-^{(i)})\right]\\
\cdot&\sin\left[\frac{\pi m_y}{d^{(i)}_{y}}(y-y_-^{(i)})\right]\cosh\left[jk^{(i)}_{z,d}(z+h^{(i)})\right]\\
H^{\{e,m\},(i)}_{\mathrm{grv},z}&=\sum_{m_x^{*}=0}^{\infty}\sum_{m_y^{*}=0}^{\infty}D^{\{e,m\},(i)}_{m_xm_y}\cos\left[\frac{\pi m_x}{d^{(i)}_{x}}(x-x_-^{(i)})\right]\\
\cdot&\cos\left[\frac{\pi m_y}{d^{(i)}_{y}}(y-y_-^{(i)})\right]\sinh\left[jk^{(i)}_{z,d}(z+h^{(i)})\right]
\end{aligned}
\end{equation}
where the $*$ symbol means that $m_x$ and $m_y$ must not vanish simultaneously. In Eq. \eqref{eq:TM_TE_modes_holes}, the complex coefficients $C^{e}_{m_{x}m_{y}} (C^{m}_{m_{x}m_{y}})$ and $D^{e}_{m_{x}m_{y}} (D^{m}_{m_{x}m_{y}})$ are the $\left(m_x,m_y\right)$-order modal weights corresponding to the TE (TM) incident field excitation scenario, and 
${[k^{(i)}_{z,\mathrm{grv}}]^{2}=k^{2}\varepsilon_{d}^{(i)}-\left[\pi m_x/d^{(i)}_{x}\right]^{2}-\left[\pi m_y/d^{(i)}_{y}\right]^{2}}$ is the modal propagation constant (longitudinal wavenumber). Although we are considering all-metallic configurations in this paper, for completeness, we provide a general derivation of the scattering problem, in which each of the grooves can be filled with a dielectric material whose relative permittivity is denoted by $\varepsilon^{(i)}_{d}$ for the $i$th groove ($\eta^{(i)}_{d}=\eta/\sqrt{\varepsilon_d^{(i)}}$ is the associated wave impedance). Eventually, for obtaining the MG designs presented in Section \ref{sec:results} herein, one should substitute $\varepsilon^{(i)}_{d}=1$ in the various expressions, indicating that the grooves are filled with vacuum.
The other (tangential) components of the electric and magnetic fields, both for the reflected fields [Eq. \eqref{eq:E_H_ref_z}] and for the fields in the grooves [Eq. \eqref{eq:TM_TE_modes_holes}], can be found by substituting the longitudinal ones into Maxwell's equations \cite{balanis2012advanced}.

\subsection{Mode matching}
\label{subsec:mode_matching}
At this point, to enable practical calculations, it is essential to truncate the infinite summations in Eqs. \eqref{eq:E_H_ref_z}-\eqref{eq:TM_TE_modes_holes}, while keeping track of the number of unknowns to ensure the eventual formation of a solvable set of equations. 
For the reflected fields [Eq. \eqref{eq:E_H_ref_z}], we truncate the sums at $\pm N_x/2$ and $\pm N_y/2$, retaining $\left(N_x+1\right)\times\left(N_y+1\right)$ FB harmonics overall. 
Accordingly, Eq. \eqref{eq:E_H_ref_z} features $2\times\left(N_x+1\right)\times\left(N_y+1\right)$ unknowns for a given polarized excitation, corresponding to the various $A^{\{e,m\}}_{n_{x}n_{y}}$ and $B^{\{e,m\}}_{n_{x}n_{y}}$. 
Similarly, for the fields inside the grooves, we will retain modes up to the order $\left(M_x, M_y\right)$. Considering Eq. \eqref{eq:TM_TE_modes_holes}, for a given polarized excitation, this corresponds to $M_x\times M_y$ unknowns per groove as per $C^{\{e,m\},\{i\}}_{n_{x}n_{y}}$ and $\left(M_x+1\right)\times\left(M_y+1\right)-1$ unknowns per groove as per $D^{\{e,m\},\{i\}}_{n_{x}n_{y}}$. Altogether, thus, Eqs. \eqref{eq:E_H_ref_z} and \eqref{eq:TM_TE_modes_holes} feature ${U\triangleq2\left(N_x+1\right)\left(N_y+1\right)+\left(2M_xM_y +M_x+M_y\right) N_{\mathrm{grv}}}$ unknowns.
For compactness, we will denote from now on the $\{n_x,n_y\}$ and $\{m_x,m_y\}$ mode indices as vectors, namely, $\bm{n}$ and $\bm{m}$ respectively; similarly, the limits of the double summations will be symbolically denoted by $\bm{N}=\{N_x,N_y\}$ and $\bm{M}=\{M_x,M_y\}$, where it is understood that the formula interpretation should follow the detailed notations in Eqs. \eqref{eq:E_H_ref_z} and \eqref{eq:TM_TE_modes_holes}.

Using the truncated sums, we enforce the boundary conditions at the interface ${z=0}$. This involves requiring the continuity of the tangential electric and magnetic field components on this plane. Defining the set of points on the $i$th groove aperture as ${\Omega^{(i)}\triangleq\left\{\left(x,y\right)|x\in\left(x_{-}^{(i)},x_{+}^{(i)}\right)\wedge y\in\left(y_{-}^{(i)},y_{+}^{(i)}\right)\right\}}$, we can formulate these continuity conditions for $E_x(x,y,0)$ and $E_y(x,y,0)$ 
on the entire unit cell area $L_{x} \times L_{y}$, leading, respectively, to 
\begin{widetext}
\begin{equation}
\label{eq:Boundary_conditions_Ex}
\begin{aligned}
\sum_{\bm{n}=-\bm{N}/2}^{\bm{N}/2}\!\!\!\!\!\!&\big[\alpha_{\bm{n}}^{(1)}A^{\{e,m\}}_{\bm{n}}\!+\!\beta_{\bm{n}}^{(1)}B^{\{e,m\}}_{\bm{n}}\!+\!s_{\bm{n}}^{\{e,m\},(1)}\delta_{\bm{n},\bm{0}}\big]e^{-jk_x x}e^{-jk_{y}y} \\ 
	&=\begin{cases}
   \sum_{\bm{m}=\bm{0}}^{\bm{M}}\big[\Gamma_{\bm{m}}^{(1,i)}C^{\{e,m\},(i)}_{\bm{m}}\!+\!\Delta_{\bm{m}}^{(1,i)}D^{\{e,m\},(i)}_{\bm{m}}\big]
   \,\cdot\cos\!\left[\frac{\pi m_x}{d^{(i)}_{x}}(x-x_{-}^{(i)})\right]\!\cdot\sin\!\left[\frac{\pi m_y}{d^{(i)}_{y}}(y-y_{-}^{(i)})\right]& \left(x,y\right)\!\in\!\Omega^{(i)}\\
    0       & \text{otherwise}
\end{cases}\\
\end{aligned}
\end{equation}
\begin{equation}
\label{eq:Boundary_conditions_Ey}
\begin{aligned}
\sum_{\bm{n}=-\bm{N/2}}^{\bm{N/2}}\!\!\!\!\!\!&\big[\alpha_{\bm{n}}^{(2)}A^{\{e,m\}}_{\bm{n}}\!+\!\beta_{\bm{n}}^{(2)}B^{\{e,m\}}_{\bm{n}}\!+\!s_{\bm{n}}^{\{e,m\},(2)}\delta_{\bm{n},\bm{0}}\big]e^{-jk_x x}e^{-jk_{y}y} \\ 
&=\begin{cases}
   \sum_{\bm{m}=\bm{0}}^{\bm{M}}\big[\Gamma_{\bm{m}}^{(2,i)}C^{\{e,m\},(i)}_{\bm{m}}\!+\!\delta_{\bm{m}}^{(2,i)}D^{\{e,m\},(i)}_{\bm{m}}\big]
   \,\cdot\sin\!\left[\frac{\pi m_x}{d^{(i)}_{x}}(x-x_{-}^{(i)})\right]\!\cdot \cos\!\left[\frac{\pi m_y}{d^{(i)}_{y}}(y-y_{-}^{(i)})\right]& \left(x,y\right)\!\in\!\Omega^{(i)}\\
    0       & \text{otherwise}
\end{cases}\\
\end{aligned}
\end{equation}
\end{widetext}
where the notations $\alpha_{\bm{n}}^{(p)},\beta_{\bm{n}}^{(p)}$ $\Gamma_{\bm{m}}^{(p,i)},\Delta_{\bm{m}}^{(p,i)}$, and $s_{\bm{n}}^{\{e,m\},(p)}$ are introduced herein for brevity, and emerge once the tangential fields are evaluated at the aperture $z=0$. For completeness, these coefficients, dependent on the modal wavenumbers ($\bm{n}$th FB mode for $\alpha_{\bm{n}}^{(p)},\beta_{\bm{n}}^{(p)}$ and $\bm{m}$th rectangular waveguide mode in the $i$th groove for $\Gamma_{\bm{m}}^{(p,i)},\Delta_{\bm{m}}^{(p,i)}$), eigenfunction properties, and the excitation polarization ($s_{\bm{n}}^{\{e,m\},(p)}$) are explicitly provided in the Appendix; the superscript $(p)$ is used merely to associate the various coefficients with the relevant boundary condition (in the order they are introduced herein). As usual, $\delta_{\bm{n},\bm{0}}$ stands for the Kronecker Delta function, which equals $1$ if $(n_x,n_y)=(0,0)$ and $0$ otherwise.


For the tangential magnetic fields, formulation of the continuity conditions on the plane $z=0$ is possible only in regions where groove openings occur, as on the perfectly conducting metal, the magnetic field is generally discontinuous due to induced surface currents. Correspondingly, for points on the $i$th groove aperture $\left(x,y\right)\in\Omega^{(i)}$ we can impose the continuity of $H_x(x,y,0)$ and $H_y(x,y,0)$, respectively, via
\begin{widetext}
\begin{equation}
\label{eq:Boundary_conditions_Hx}
\begin{aligned}
\sum_{\bm{n}=-\bm{N}/2}^{\bm{N}/2}\!\!\!\!&\big[\alpha_{\bm{n}}^{(3)}A^{\{e,m\}}_{\bm{n}}\!+\!\beta_{\bm{n}}^{(3)}B^{\{e,m\}}_{\bm{n}}\!+\!s_{\bm{n}}^{\{e,m\},(3)}\delta_{\bm{n},\bm{0}}\big]e^{-jk_x x}e^{-jk_{y}y}\\
&=\sum_{\bm{m}=\bm{0}}^{\bm{M}}\big[\Gamma_{\bm{m}}^{(3,i)}C^{\{e,m\},(i)}_{\bm{m}}+\Delta_{\bm{m}}^{(3,i)}D^{\{e,m\},(i)}_{\bm{m}}\big]
\cdot \sin\left[\frac{\pi m_x}{d^{(i)}_{x}}(x-x_{-}^{(i)})\right]\cdot \cos\left[\frac{\pi m_y}{d^{(i)}_{y}}(y-y_{-}^{(i)})\right]
\end{aligned}
\end{equation}
\begin{equation}
\label{eq:Boundary_conditions_Hy}
\begin{aligned}
\sum_{\bm{n}=-\bm{N}/2}^{\bm{N}/2}\!\!\!\!&\big[\alpha_{\bm{n}}^{(4)}A^{\{e,m\}}_{\bm{n}}\!+\!\beta_{\bm{n}}^{(4)}B^{\{e,m\}}_{\bm{n}}\!+\!s_{\bm{n}}^{\{e,m\},(4)}\delta_{\bm{n},\bm{0}}\big]e^{-jk_x x}e^{-jk_{y}y}
\\
&=\sum_{\bm{m}=\bm{0}}^{\bm{M}}\big[\Gamma_{\bm{m}}^{(4,i)}C^{\{e,m\},(i)}_{\bm{m}}+\Delta_{\bm{m}}^{(4,i)}D^{\{e,m\},(i)}_{\bm{m}}\big]
\cdot \cos\left[\frac{\pi m_x}{d^{(i)}_{x}}(x-x_{-}^{(i)})\right]\cdot \sin\left[\frac{\pi m_y}{d^{(i)}_{y}}(y-y_{-}^{(i)})\right]
\end{aligned}
\end{equation}
\end{widetext}
where, again, the expressions for the coefficients $\alpha_{\bm{n}}^{(p)},\beta_{\bm{n}}^{(p)}$ and $\Gamma_{\bm{m}}^{(p,i)},\Delta_{\bm{m}}^{(p,i)}$ for the third [Eq. \eqref{eq:Boundary_conditions_Hx}] and fourth boundary conditions [Eq. \eqref{eq:Boundary_conditions_Hy}] can be found in the Appendix.

We resolve the conditions in Eqs. \eqref{eq:Boundary_conditions_Ex}-\eqref{eq:Boundary_conditions_Hy} via mode-matching, harnessing the orthogonality of the FB harmonics in free space and the guided modes in the grooves \cite{schachter1997periodic}. Specifically, for every combination $(n'_x, n'_y)$ of FB mode indices, we multiply Eqs. \eqref{eq:Boundary_conditions_Ex} and \eqref{eq:Boundary_conditions_Ey} by $e^{jk^{(n'_x)}x}\cdot e^{jk^{(n'_y)}y}$ and integrate over the period $L_x\times L_y$, eventually arriving, respectively, at \footnote{In order to avoid cumbersome notations, once the integration is completed, we perform a final change of variables, denoting $\mathbf{n}=\left(n'_x, n'_y\right)$.} 
\begin{equation}
\label{eq:Final_equation_Ex}
\begin{aligned}
&\alpha_{\bm{n}}^{(1)}A^{\{e,m\}}_{\bm{n}}+\beta_{\bm{n}}^{(1)}B^{\{e,m\}}_{\bm{n}}
\\
&-\sum_{i=1}^{N_\mathrm{grv}}\sum_{\bm{m}=\bm{0}}^{\bm{M}}\big[\Gamma_{\bm{m}}^{(1,i)}C^{\{e,m\},(i)}_{\bm{m}}+\Delta_{\bm{m}}^{(1,i)}D^{\{e,m\},(i)}_{\bm{m}}\big]\psi^{(i)}_{\bm{n,m}}\\
&=-S_{\bm{n}}^{(1)}
\end{aligned}
\end{equation}
and
\begin{equation}
\label{eq:Final_equation_Ey}
\begin{aligned}
&\alpha_{\bm{n}}^{(2)}A^{\{e,m\}}_{\bm{n}}+\beta_{\bm{n}}^{(2)}B^{\{e,m\}}_{\bm{n}}
\\
&-\sum_{i=1}^{N_\mathrm{grv}}\sum_{\bm{m}=\bm{0}}^{\bm{M}}\big[\Gamma_{\bm{m}}^{(2,i)}C^{\{e,m\},(i)}_{\bm{m}}+\Delta_{\bm{m}}^{(2,i)}D^{\{e,m\},(i)}_{\bm{m}}\big]\chi^{(i)}_{\bm{n,m}}\\
&=-S_{\bm{n}}^{(2)}
\end{aligned}
\end{equation}
In Eqs. \eqref{eq:Final_equation_Ex} and  \eqref{eq:Final_equation_Ey}, $\psi^{(i)}_{\bm{n},\bm{m}}$ and $\chi^{(i)}_{\bm{n},\bm{m}}$ are the overlap integrals between the $(n_x,n_y)$-order FB mode and the $(m_x,m_y)$-order guided mode on the aperture of the $i$th groove, given by
\begin{equation}
\label{eq:psi}
\begin{aligned}
\psi^{(i)}_{\bm{n},\bm{m}} = &\frac{1}{L_x L_y}\int_{x^{(i)}_-}^{x^{(i)}_+}\int_{y^{(i)}_-}^{y^{(i)}_+}e^{jk^{(n_x)}_xx}e^{jk^{(n_y)}_yy} \\
&\cdot \cos\bigg[\frac{\pi m_x}{d^{(i)}_x}(x-x^{(i)}_-)\bigg]\sin\bigg[\frac{\pi m_y}{d^{(i)}_y}(y-y^{(i)}_-)\bigg]dxdy
\end{aligned}
\end{equation}
and
\begin{equation}
\label{eq:chi}
\begin{aligned}
\chi^{(i)}_{\bm{n},\bm{m}} = &\frac{1}{L_x L_y}\int_{x^{(i)}_-}^{x^{(i)}_+}\int_{y^{(i)}_-}^{y^{(i)}_+}e^{jk^{(n_x)}_xx}e^{jk^{(n_y)}_yy} \\
&\cdot \sin\bigg[\frac{\pi m_x}{d^{(i)}_x}(x-x^{(i)}_-)\bigg]\cos\bigg[\frac{\pi m_y}{d^{(i)}_y}(y-y^{(i)}_-)\bigg]dxdy
\end{aligned}
\end{equation}
which can be evaluated analytically and written in closed form, as laid out in Eq. \eqref{eq:Analytical_exp} in the Appendix. The term $S_{\bm{n}}^{(p)}$ in Eqs. \eqref{eq:Final_equation_Ex} and \eqref{eq:Final_equation_Ey} is associated with the TE or TM excitation source and is given in the Appendix as well.


For resolving Eqs. \eqref{eq:Boundary_conditions_Hx} and \eqref{eq:Boundary_conditions_Hy}, defined for the points $\left(x,y\right)\in\Omega^{(i)}$, the orthogonality of the $i$th groove eigenmodes can be utilized in a similar manner. In particular, for every groove $i$ and combination $(m'_x,m'_y)$ of guided mode indices ($m'_x\neq 0$), we multiply Eq. \eqref{eq:Boundary_conditions_Hx} by $\sin[(\pi m'_{x}/d_x^{(i)})(x-x_{-}^{(i)})]\cos[(\pi m'_{y}/d_y^{(i)})(y-y_{-}^{(i)})]$ and integrate over the aperture $\Omega^{(i)}$, leading to \footnote{Similar to Eqs. \eqref{eq:Final_equation_Ex} and \eqref{eq:Final_equation_Ey}, here as well we eventually change variables and denote $\mathbf{m}=\left(m'_x, m'_y\right)$.}
\begin{equation}
\label{eq:Final_equation_Hx}
\begin{aligned}
&\sum_{\bm{n}=-\bm{N}/2}^{\bm{N}/2}\big[\alpha_{\bm{n}}^{(3)}A^{\{e,m\}}_{\bm{n}}+\beta_{\bm{n}}^{(3)}B^{\{e,m\}}_{\bm{n}}\big][\chi^{(i)}]^{*}_{\bm{n,m}}
\\
&-\big[\Gamma_{\bm{m}}^{(3,i)}C^{\{e,m\},(i)}_{\bm{m}}+\Delta_{\bm{m}}^{(3,i)}D^{\{e,m\},(i)}_{\bm{m}}\big]=-S_{\bm{m}}^{(3)}
\end{aligned}
\end{equation}
In a dual manner, multiplying Eq. \eqref{eq:Boundary_conditions_Hy} by 
$\cos[(\pi m'_x/d_x^{(i)})(x-x_{-}^{(i)})]\sin[(\pi m'_y/d_y^{(i)})(y-y_{-}^{(i)})]$ for $m'_y\neq 0$, and integrating as in Eq. \eqref{eq:Final_equation_Hx}, yields another set of equations, reading 
\begin{equation}
\label{eq:Final_equation_Hy}
\begin{aligned}
&\sum_{\bm{n}=-\bm{N/2}}^{\bm{N/2}}\big[\alpha_{\bm{n}}^{(4)}A^{\{e,m\}}_{\bm{n}}+\beta_{\bm{n}}^{(4)}B^{\{e,m\}}_{\bm{n}}\big][\psi^{(i)}]^{*}_{\bm{n,m}}
\\
&-\big[\Gamma_{\bm{m}}^{(4,i)}C^{\{e,m\},(i)}_{\bm{m}}+\Delta_{\bm{m}}^{(4,i)}D^{\{e,m\},(i)}_{\bm{m}}\big]=-S_{\bm{m}}^{(4)}
\end{aligned}
\end{equation}

Let us now review the set of equations we have arrived at after applying the inner product operations above, recalling that the parameters $\alpha_{\bm{n}}^{(p)},\beta_{\bm{n}}^{(p)}$ and $\Gamma_{\bm{m}}^{(p,i)},\Delta_{\bm{m}}^{(p,i)}$ are known for a given MG groove geometry (Appendix). Hence, Eqs. \eqref{eq:Final_equation_Ex}-\eqref{eq:Final_equation_Ey} each formulate $N_x\times N_y$ relations between the unknown scattering coefficients $A^{\{e,m\}}_{n_{x}n_{y}}$, $B^{\{e,m\}}_{n_{x}n_{y}}$, $C^{\{e,m\}}_{m_{x}m_{y}}$, and $D^{\{e,m\}}_{m_{x}m_{y}}$; overall, they form $2\times N_x\times N_y$ such linear equations. Furthermore, Eq. \eqref{eq:Final_equation_Hx} contributes additional $M_x\times M_y + M_y$ equations per groove ($m_x\neq 0$) and Eq. \eqref{eq:Final_equation_Hy} contributes $M_x\times M_y + M_x$ equations per groove ($m_y\neq 0$), yielding altogether another $\left(2\times M_x\times M_y + M_x+ M_y\right)\cdot N_{\mathrm{grv}}$ relations for the scattering coefficients.
Thus, in total, Eqs. \eqref{eq:Final_equation_Ex}, \eqref{eq:Final_equation_Ey}, \eqref{eq:Final_equation_Hx}, and \eqref{eq:Final_equation_Hy} form $U$ linear equations with $U$ unknowns, enabling solution of the problem via a simple matrix inversion.

Indeed, the set of linear constraints can be cast in the form of a matrix equation, 
\begin{equation}
\label{eq:Matrix_equation}
\begin{aligned}
\begin{pmatrix}
  \mathbf{Z_{U\times U}}   \vspace{3pt}
  \end{pmatrix}
  \begin{pmatrix}
  \mathbf{I_{U\times 1}}
  \end{pmatrix}
  =  
  \begin{pmatrix}
  \mathbf{V_{U\times 1}}   \vspace{3pt}
  \end{pmatrix}
\end{aligned}
\end{equation}
where the impedance matrix elements of $\mathbf{Z_{U\times U}}$ are formed by the coefficients $\alpha^{(p)}_{\bm{n}},\beta^{(p)}_{\bm{n}},\Gamma^{(p)}_{\bm{m}},\Delta^{(p)}_{\bm{m}}$ ($p=1,2,3,4$), with each matrix row corresponding to one of the equations \eqref{eq:Final_equation_Ex}, \eqref{eq:Final_equation_Ey}, \eqref{eq:Final_equation_Hx}, and \eqref{eq:Final_equation_Hy}.
The current vector $\mathbf{I_{U\times 1}}$ consists of the unknowns $A^{\{e,m\}}_{\bm{n}},B^{\{e,m\}}_{\bm{n}},C^{\{e,m\}}_{\bm{m}},D^{\{e,m\}}_{\bm{m}}$, and the source vector $\mathbf{V_{U\times 1}}$ is associated with the TM- or TE- polarized plane wave excitation terms, corresponding to $S_{\bm{n}}^{(1)}$, $S_{\bm{n}}^{(2)}$, $S_{\bm{m}}^{(3)}$, and $S_{\bm{m}}^{(4)}$ as defined in these equations (see Appendix). 
For practical calculations made in this work (Section \ref{sec:results}), we have found that using $M_x=M_y=5$ and $N_x=N_y=10$ to truncate the sums of Eqs. \eqref{eq:E_H_ref_z}-\eqref{eq:TM_TE_modes_holes} 
was sufficient for the convergence of the solution.

\subsection{Perfect anomalous reflection}
\label{subsec:anomalous_reflection}
Once we established Eq. \eqref{eq:Matrix_equation}, we can readily retrieve the scattered fields for a given MG configuration. Therefore, the next step in the MG synthesis procedure would be to formulate constraints on these scattering coefficients such that the desirable functionality - in our case, co-polarized perfect anomalous reflection for either TE or TM (or both) incident fields - will be implemented by the devised device. Although the formalism allows, in principle, quite flexible control of the power partition to the various $N_{\mathrm{prop}}$ reflected propagating FB modes, we focus herein on a specific prototypical application, requiring that all the incident power of a given polarization will be funnelled to the $(n_x, 0)$-order FB mode having the same polarization [Fig. \ref{Fig:Configuration}(b)]. 

Dividing the anomalously reflected power by the incident power using Eqs. (\ref{eq:E_H_inc}), (\ref{eq:E_H_ref_z}) and the retrieved $A^{m}_{n_{x},0}$ and $B^{e}_{n_{x},0}$ coefficients yields, respectively, the $(n_x,0)$ anomalous reflection efficiency for TE and TM excitations, defined as
\begin{equation}
\label{eq:eta_condition}
\begin{aligned}
&\eta^{e}_{n_x,0}=\frac{|B_{n_x,0}^{e}|^{2} k\cdot k_{z}}{|E^{e}_{0}/\eta|^{2}\cos\theta_{\mathrm{in}} [k^{(n_x)}_{x}]^{2}}\\
&\eta^{m}_{n_x,0}=\frac{|A_{n_x,0}^{m}|^{2} k\cdot k_{z}}{|H^{m}_{0}|^{2}\cos\theta_{\mathrm{in}} [k^{(n_x)}_{x}]^{2}}
\end{aligned}
\end{equation}
Seeking optimal performance, we desire that $\eta^{e}_{n_x,0}\to 1$ or $\eta^{m}_{n_x,0}\to 1$ (or both), depending on desired polarization response, which actually implies that the coupling to the other $N_{\mathrm{prop}}-1$ propagating FB modes should be suppressed. This, in turn, translates into $N_{\mathrm{prop}}-1$ nonlinear constraints (per polarization),  
which can be solved, in principle, by using at least $N_{\mathrm{prop}}-1$ DOFs, corresponding to the MG geometrical parameters [Fig. \ref{Fig:Configuration}(a)].

This step completes the synthesis procedure. Correspondingly, to design an all-metallic MG as in Fig. \ref{Fig:Configuration} that would couple the incident TE- or TM- polarized power incoming from $\theta_\mathrm{in}$ towards $\theta_\mathrm{out}$ in its entirety, we should solve the matrix equation in Eq. \eqref{eq:Matrix_equation}, under the corresponding nonlinear set of constraints that guarantee that all the undesired radiation channels would vanish as per Eq. \eqref{eq:eta_condition}.
The solution to these nonlinear equations can be achieved graphically (as in Sections \ref{subsec:singl_pol_one_groove} and \ref{subsec:dual_pol}) or using a standard library function in MATLAB (as in Section \ref{subsec:singl_pol_two_grooves}), eventually yielding a detailed groove configuration, suitable for fabrication via computer numerical control (CNC) machining, which implements the required beam deflection.

\section{Results and Discussion}
\label{sec:results}
\subsection{Single-polarized anomalous reflection\\(two radiation channels)}
\label{subsec:singl_pol_one_groove}
\begin{figure}[t]
\centering
\includegraphics[width=3.2in]{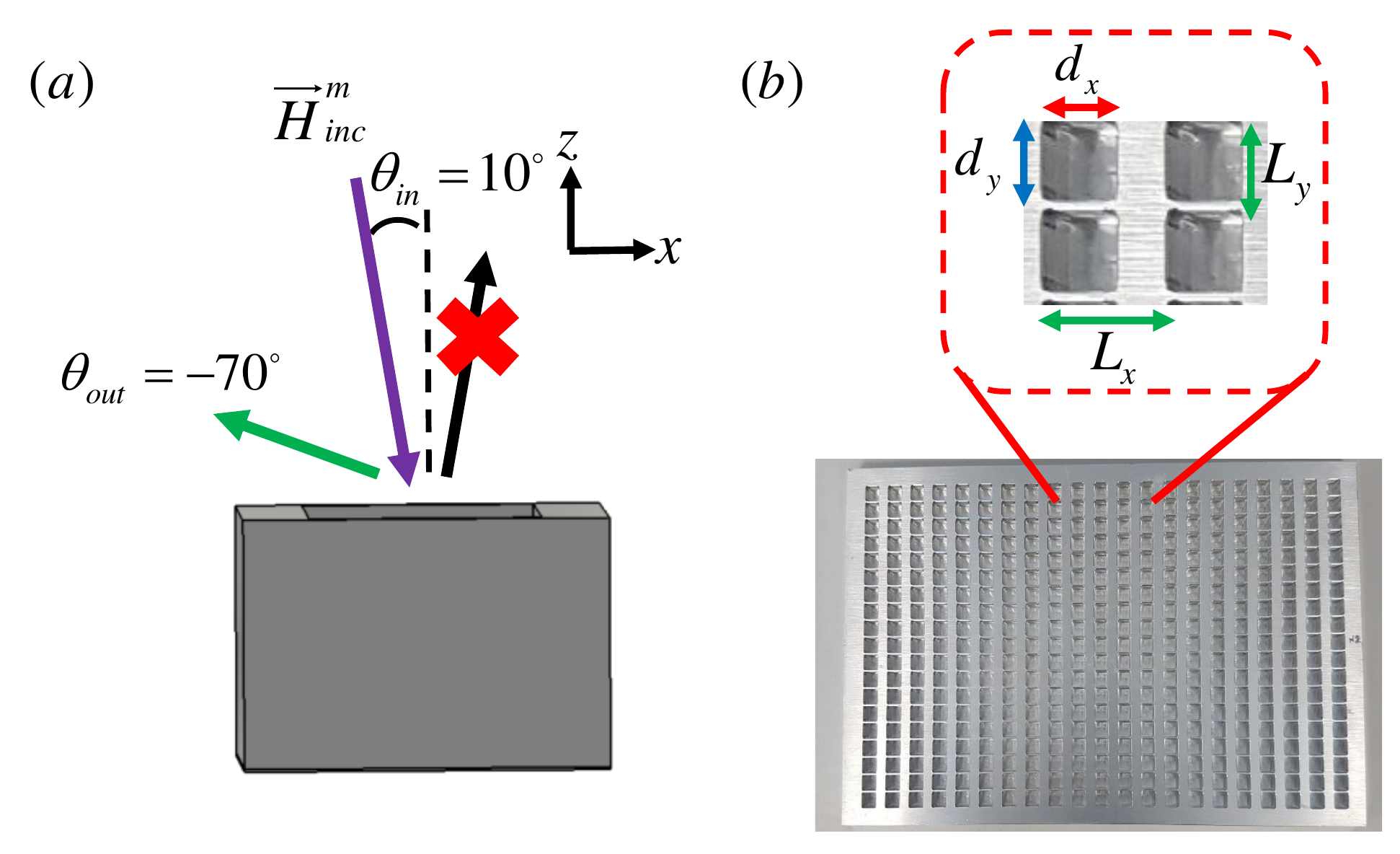}
\caption{MG designed for TM-polarized anomalous reflection from $\theta_{\mathrm{in}}=10^{\circ}$ towards $\theta_{\mathrm{out}}=-70^{\circ}$ at $f=20$ GHz, featuring two propagating FB modes. (a) Physical configuration (one groove per period). 
(b) Manufactured prototype. Inset: close-up on 4 unit cells, showing the $1$ mm curvature radius at the groove corners, stemming from CNC fabrication limitations.}
\label{Fig:Config_10tominus70}
\end{figure}

To verify the theoretical derivation and demonstrate its applicability, we utilize the analytical formalism to design several prototypical MGs for anomalous reflection (of increasing complexity), fabricate them, and characterize them experimentally. We start with a basic anomalous reflection functionality, targeting a single-polarization MG (at $f=20$ GHz) that would redirect all the power carried by a TM-polarized plane wave with angle of incidence $\theta_\mathrm{in}=10^{\circ}$ towards $\theta_\mathrm{out}=-70^{\circ}$ (Fig. \ref{Fig:Config_10tominus70}). For these angles of incidence and deflection, the required MG periodicity in the $x$ direction $L_{x}=\lambda/|\sin\theta_{\mathrm{in}}-\sin\theta_{\mathrm{out}}|=13.47$ mm $\approx 0.9\lambda$ dictates that only two FB modes will be propagating, and the rest will be evanescent, as long as the periodicity along the $y$ axis satisfies $L_{y}<\lambda\approx15$ mm \cite{tretyakov2003analytical,rabinovich2018analytical}. Thus, in this case, there are only \textit{two} reflection channels relevant for far-field radiation: the specular [the $(n_{x},n_{y})=(0,0)$ FB mode] and the anomalous [the $(n_{x},n_{y})=(-1,0)$ harmonic]. To exclusively couple all the incident power to the $(-1,0)$ harmonic, suppression of a \textit{single} reflection channel is required.

Since we only need to satisfy a single constraint, one DOF should, in principle, suffice. Therefore, a single groove per period can be used in this case [Fig. \ref{Fig:Config_10tominus70}(a)]. As even this very basic configuration features multiple geometrical parameters [Fig. \ref{Fig:Configuration}(a)], we retain only two of them as DOFs, namely, the depth of the groove $h$ and its width (along the $x$ axis) $d_x$, and fix the rest as $L_y=10$ mm and $d_y=0.9L_y=9$ mm (other parameter values can be used as well, if desired); without loss of generality, the center of the groove is chosen as $(a_{x},a_{y})=(0,0)$.

\begin{figure}[t]
\centering
\includegraphics[width=2.5in]{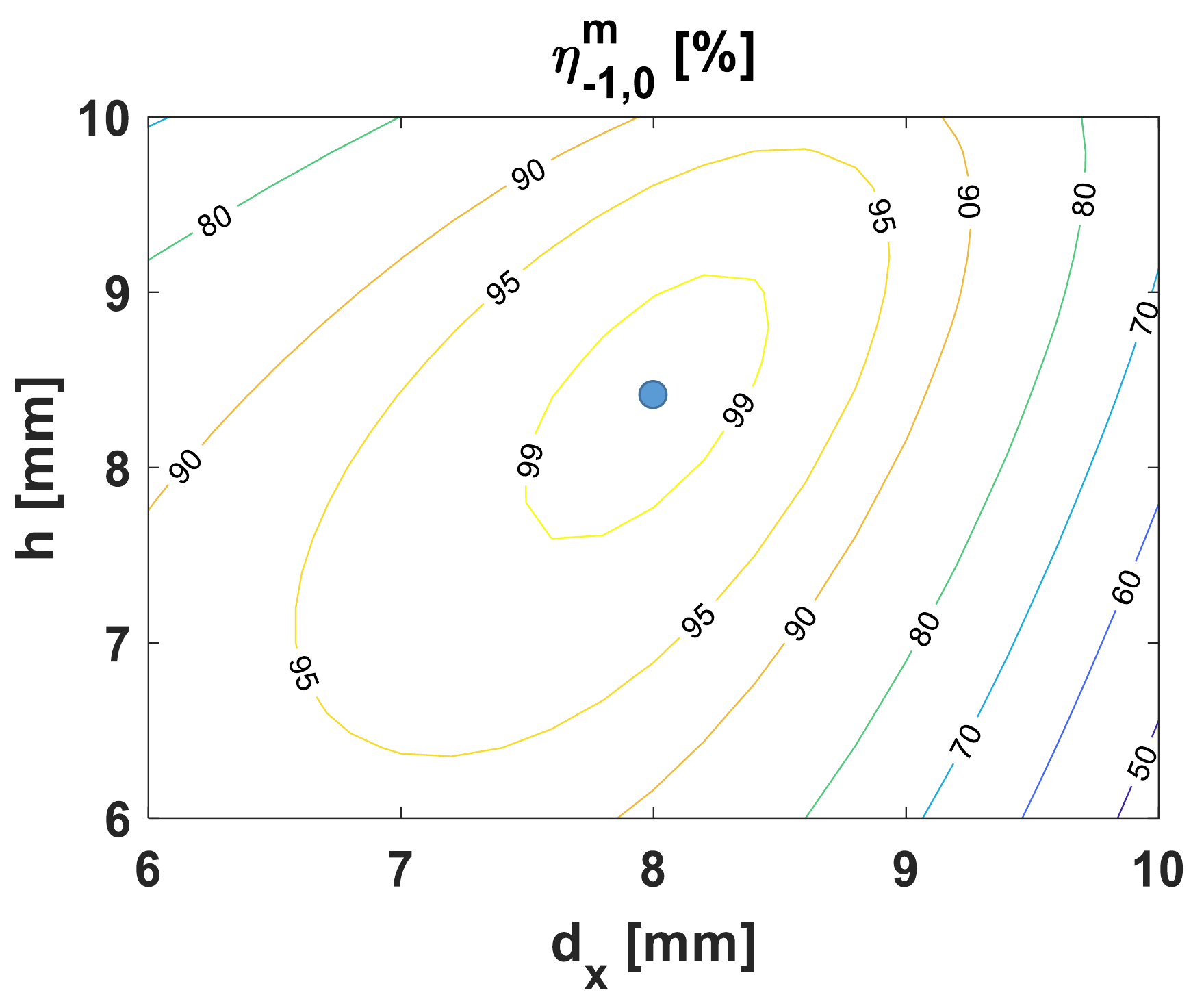}
\caption{Anomalous reflection efficiency $\eta^{m}_{-1,0}$ for the single-polarized anomalous reflection scenario corresponding to Fig. \ref{Fig:Config_10tominus70}(a), as function of the groove depth $h$ and the width $d_x$ [Eqs. \eqref{eq:Matrix_equation}, \eqref{eq:eta_condition}]. The optimal working point eventually selected for the prototype design, ($d_x=8$ mm, $h=8.4$ mm), is marked with a blue circle.}
\label{Fig:eta_10to_minus70_TM}
\end{figure}

Once the suitable number of DOFs is identified, we set to apply the previously formulated constraints as per Eq. \eqref{eq:eta_condition} to guarantee TM-polarized anomalous reflection, demanding that $\eta^{m}_{-1,0}\to 1$.
For the small number of DOFs considered in this case, a simple sweep of the geometrical parameters can be readily performed using the analytical model [Eq. \eqref{eq:Matrix_equation}], which is convenient for observing general trends and obtaining an optimal solution.
Figure \ref{Fig:eta_10to_minus70_TM} presents a contour 2D-plot of the anomalous reflection efficiency $\eta^{m}_{-1,0}$ as a function of $d_x$ and $h$, in percentages. As observed, the plot reveals a region of possible combinations $\left(d_x,h\right)$ that would enable high anomalous reflection efficiency ($>95\%$). For our prototype design, we choose the geometrical parameters leading to the highest efficiency, namely, $(d_{x},h)=(8,8.4)$ mm (marked by a blue circle in Fig. \ref{Fig:eta_10to_minus70_TM}), for which $99.9\%$ of the incident power is expected to be redirected towards $-70^\circ$. 



\begin{figure}[t]
\centering
\includegraphics[width=2.5in]{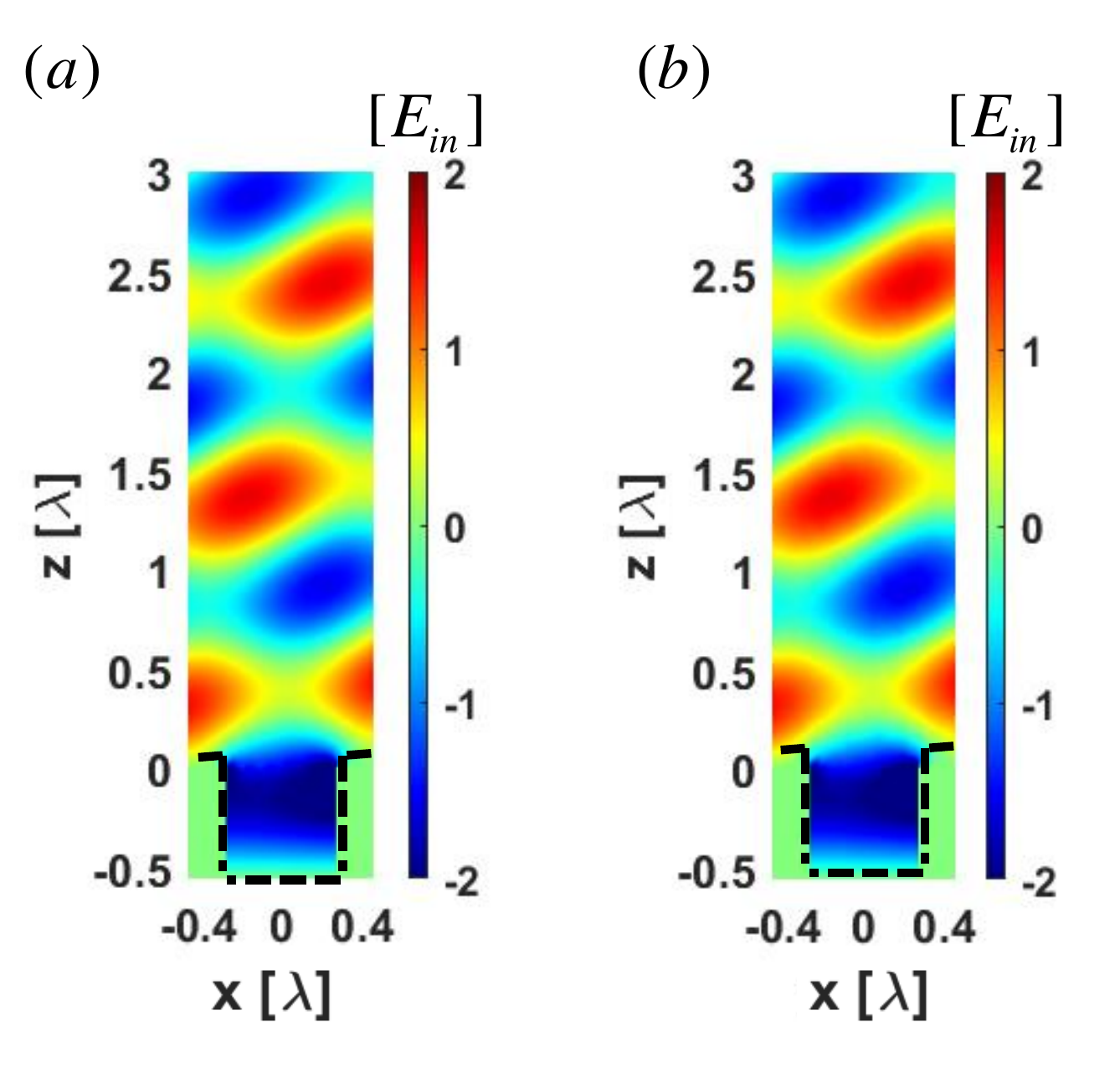}
\caption{Electric field distribution $\Re\left\{E_x(x,y=0,z)\right\}$ corresponding to the single-polarized anomalous reflector of Fig. \ref{Fig:Config_10tominus70}, when illuminated from $\theta_{\mathrm{in}}=10^{\circ}$ by a TM-polarized plane wave at $f=20$ GHz (single period is shown). 
The analytical prediction (a) following Eq. \eqref{eq:Matrix_equation} with Eqs. \eqref{eq:E_H_inc}-\eqref{eq:TM_TE_modes_holes}, is compared with full-wave simulation results (b). The black dashed lines mark the boundaries of the metallic construct.}
\label{Fig:Field_10tominus70}
\end{figure}

In order to verify the analytical calculations, we compare the scattered fields theoretically predicted by the model for the chosen MG configuration with full-wave simulations conducted using CST Microwave Studio (aluminium with realistic conductivity of $\sigma = 3.56 \times 10^{7}$ S/m was used for the metallic construct). 
Field snapshots $\Re\{E_{x}(x,y=0,z)\}$ evaluated on the $\widehat{xz}$ plane are correspondingly plotted in Fig. \ref{Fig:Field_10tominus70}, showing an excellent agreement between the analytical and full-wave results. The peak anomalous reflection efficiency recorded in simulations reaches $99.6\%$ at $f=20$ GHz, confirming the fidelity of the synthesis procedure.

After this validation using a commercial solver, a corresponding finite $9''\times12''$ (thickness $H=12.7$ mm)
MG prototype was fabricated from aluminium using CNC technology for experimental characterization. Due to the fabrication process limitations, perfectly sharp corners could not be machined, and the right angles used in the model and simulations for the groove wall geometry were deformed in reality into rounded corners with a radius of curvature of $\approx 1$ mm [see inset of Fig. \ref{Fig:Config_10tominus70}(b)]. Nevertheless, full-wave simulations of the realistic geometry indicated that this minor deformation did not significantly affect the MG performance.

\begin{figure}[t]
\centering
\includegraphics[width=3.4in]{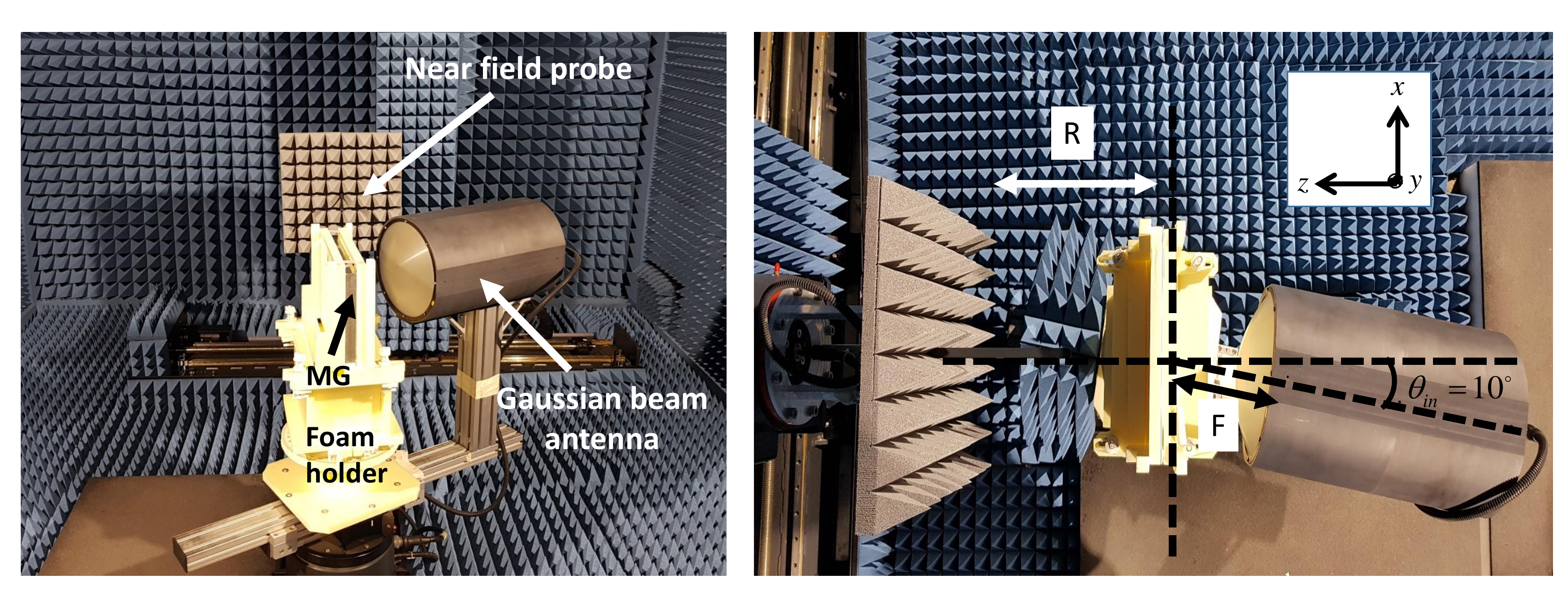}
\caption{Experimental setup used to characterize the fabricated MGs, featuring
a Gaussian beam antenna, a near field probe, and the all-metallic MG (DUT), properly positioned in an anechoic chamber in the Technion. The exciting Gaussian beam antenna and the MG are rotated together to facilitate a cylindrical near-field measurement, retaining their relative angle ($\theta_\mathrm{in}$). The distances between the near-field probe and the DUT, and between the DUT and the Gaussian beam antenna, are denoted by $R$ and $F$, respectively.}
\label{Fig:EXP_Setup}
\end{figure}

The fabricated MG was subsequently placed in an anechoic chamber in the Technion, where a cylindrical near-field measurement was conducted. The measurement setup is shown in Fig. \ref{Fig:EXP_Setup}; it is composed of a Gaussian beam antenna (Millitech, Inc., GOA-42-S000094, focal distance of $196$ mm $\approx$ 13 $\lambda$) attached to a metallic arm, a foam holder upon which the MG slab (device under test, DUT) was mounted, and an open-ended waveguide probe connected to a near-field measurement and data processing system (MVG/Orbit-FR). The DUT was placed at the focus of the Gaussian beam antenna, with the relative angle between the two fixed to be $\theta_\mathrm{in}=10^{\circ}$; the distance between the DUT and the probe was $R=800$ mm $\approx 53\lambda$. In the course of the measurement, the Gaussian beam antenna and the MG were azimuthally rotated together (retaining their relative angle), while the probe moved up and down periodically, recording the fields scattered from the MG on a cylindrical shell of radius $R$ around it. The collected data was post-processed by MiDAS data acquisition and analysis software, yielding the far-field scattering pattern. For reference,  
we also measured the radiation pattern of the Gaussian beam antenna in the absence of the MG, allowing evaluation of the overall incident power for efficiency calculations. 

\begin{figure}[t]
\centering
\includegraphics[width=2.5in]{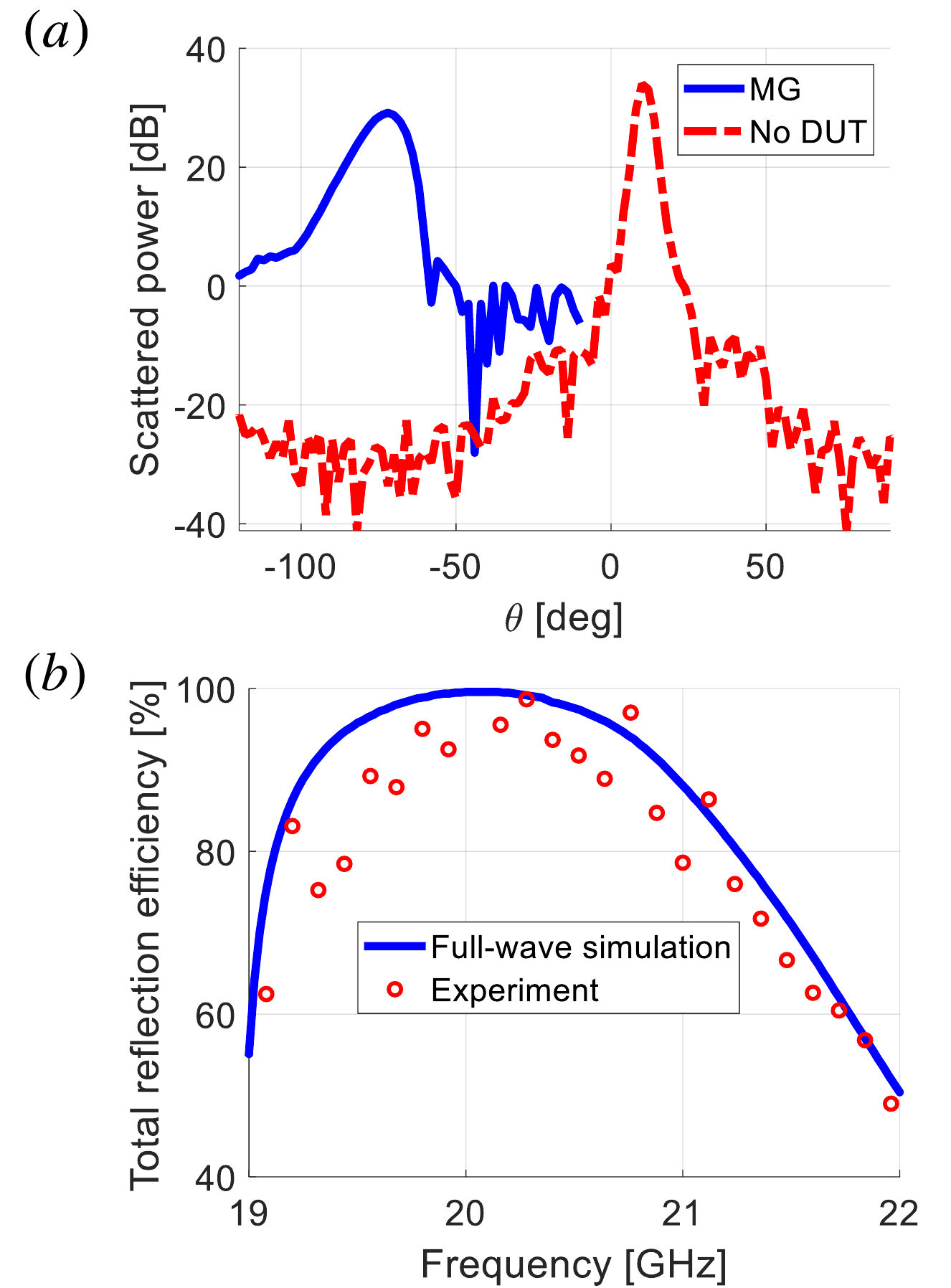}
\caption{Experimental characterization of the single-polarized anomalous reflection MG of Fig. \ref{Fig:Config_10tominus70}. (a) Received power ($f=20$ GHz) as a function of the observation angle $\theta$ with respect to the $z$ axis (scattering pattern) as obtained when the MG was excited (solid blue) is compared to the reference pattern recorded in the absence of the DUT (dash-dotted red).
(b) Total anomalous reflection efficiency $\eta^m_\mathrm{tot}$ of the MG prototype towards the $(n_x,n_y)=(-1,0)$ FB mode as a function of frequency. Experimental results (red circles) are compared with the ones obtained via full-wave simulation (solid blue).}
\label{Fig:EXP_10to_minus70_TM}
\end{figure}

The far-field scattering patterns recorded at the operating frequency $f=20$ GHz are shown
in Fig. \ref{Fig:EXP_10to_minus70_TM}(a). 
As can be clearly seen, the MG redirects the incoming power (dash-dotted red) towards the designed output angle $\theta_{\mathrm{out}}=-70^{\circ}$ (solid blue). Due to the limitations of the experimental setup, blockage effects presented by the Gaussian beam antenna prevent reliable evaluation of the reflected power by the near-field probe for certain azimuthal rotation angles. For this reason, the scattering pattern of the MG is measured over a limited angular range, and the specular reflection, supposed to be received at $\theta_{\mathrm{out}}=10^{\circ}$ is not observed in the figure. 
Nevertheless, the total anomalous reflection efficiency $\eta^m_\mathrm{tot}$, quantitatively evaluated by comparing the peak gain measured in the presence of the MG $G^m_{\mathrm{MG}}(\theta_{\mathrm{out}})$ and in its absence $G^m_{\mathrm{direct}}(\theta_{\mathrm{in}})$ as in \cite{diaz2017generalized,wong2018perfect,rabinovich2019arbitrary}

\begin{equation}
\label{eq:efficiency_EXP}
\begin{aligned}
\eta^{\{e,m\}}_{\mathrm{tot}}(f)=
\end{aligned}\frac{G^{\{e,m\}}_{\mathrm{MG}}(\theta_{\mathrm{out}})}{G^{\{e,m\}}_{\mathrm{direct}}(\theta_{\mathrm{in}})}\frac{\cos\theta_{\mathrm{in}}}{\cos\theta_{\mathrm{out}}}
\end{equation}
is not affected by this limitation, indicating effective suppression of all undesirable scattering [Fig. \ref{Fig:EXP_10to_minus70_TM}(b)]. This quantity, taking into account both absorption and spurious reflections, reaches a peak value of $98.7\%$ at $f=20.28$ GHz, validating the efficacy of the fabricated MG reflector. 
The small shift with respect to the designated operating frequency ($\sim 1.5\%$) can be attributed to difficulties in exact azimuthal alignment of the Gaussian beam antenna with respect to the characterized MG. Nonetheless, a very good correspondence between full-wave simulations and experimental results is observed overall, indicating the successful realization of a wide-angle all-metallic perfect anomalous reflection MG using the semianalytical methodology presented in Section \ref{sec:theory}.  

\subsection{Single-polarized anomalous reflection\\(three radiation channels)}
\label{subsec:singl_pol_two_grooves}
\begin{figure}[t]
\centering
\includegraphics[width=3.2in]{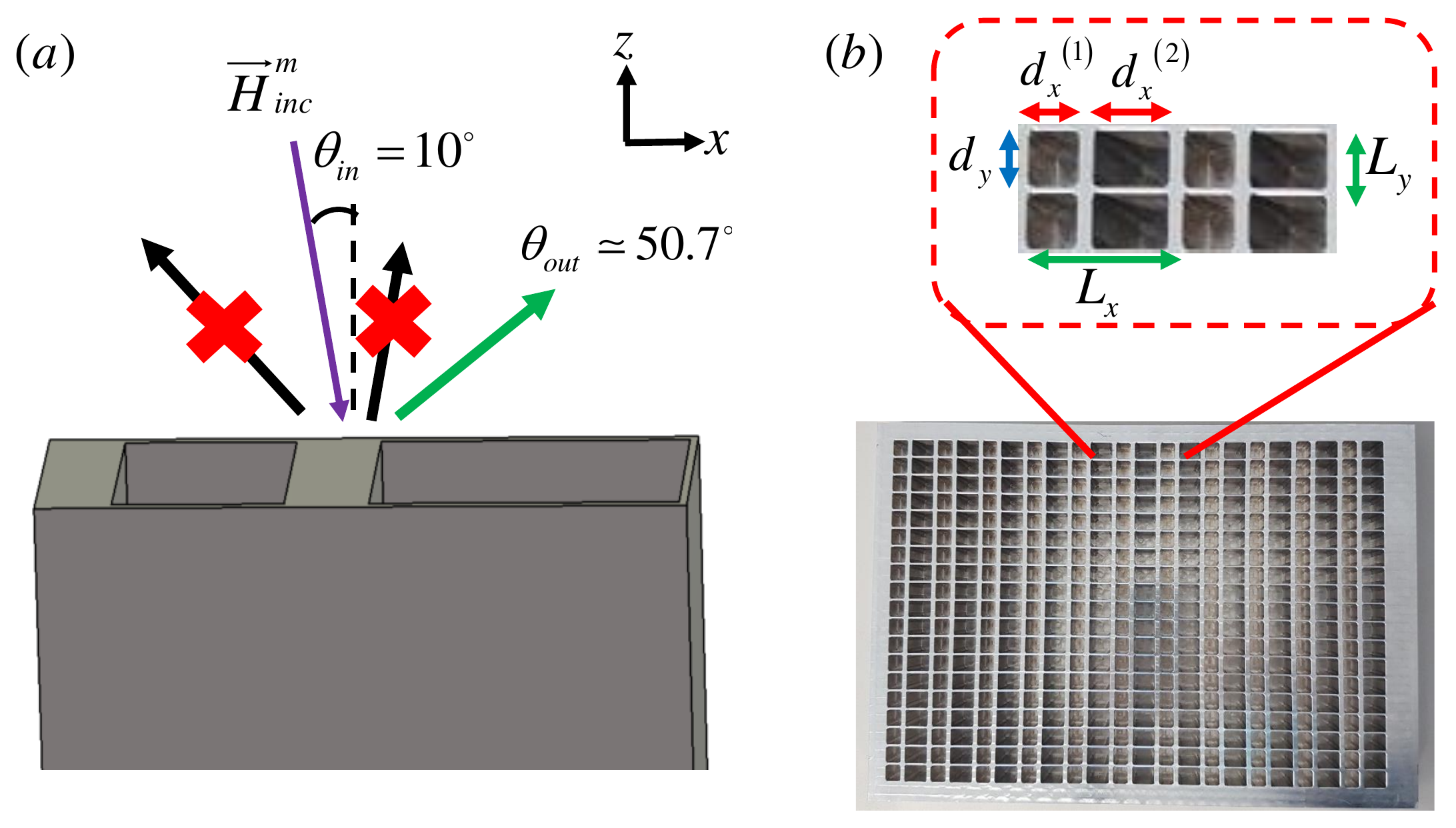}
\caption{MG designed for TM-polarized anomalous reflection from $\theta_{\mathrm{in}}=10^{\circ}$ towards $\theta_{\mathrm{out}}=50.7^{\circ}$ at $f=20$ GHz, featuring three propagating FB modes (Section \ref{subsec:singl_pol_two_grooves}). (a) Physical configuration (two grooves per period). 
(b) Manufactured prototype. Inset: close-up on 4 unit cells; the radius of curvature of the deformed groove corners, related to fabrication constraints, is $1.5$ mm in this more demanding case.}
\label{Fig:Config_10to50}
\end{figure}

To further demonstrate the versatility of the proposed semianalytical scheme, we consider next a more intricate anomalous reflection functionality, involving \emph{three} propagating FB modes potentially reflected from the MG. In particular, our goal is to design a MG that would couple all the power from the same TM-polarized incident plane wave $\theta_{\mathrm{in}}=10^{\circ}$ (at $f=20$ GHz) towards $\theta_{\mathrm{out}}=50.7^{\circ}$ (Fig. \ref{Fig:Config_10to50}). We follow the same reasoning presented in Section \ref{subsec:singl_pol_one_groove}, and choose $L_y=10$ mm. Moving the anomalous reflection angle to the first quadrant of the $\widehat{xz}$ plane implies greater periodicity $L_{x}=\lambda/|\sin\theta_{\mathrm{in}}-\sin\theta_{\mathrm{out}}|=25$ mm = $1.67\lambda$ , which would indeed open a third reflection channel, allowing the FB modes of order $(n_{x},n_{y})=(-1,0)$, $(0,0)$, and $(1,0)$ to propagate. Thus, in this case, to achieve perfect anomalous reflection, we need to suppress coupling to \emph{two} FB modes, namely, the specular reflection and the $(-1,0)$-order mode, such that all the incoming power would be funnelled to the desired $(1,0)$ anomalous reflection mode.

Considering the additional constraint in this case, and relying on the observations from Section \ref{subsec:singl_pol_one_groove}, we choose to utilize herein a MG featuring two grooves per period, to correspondingly increase the number of available DOFs. Once again, we keep as DOFs the width and depth of the two grooves, namely, $d_{x}^{(1)}$, $d_{x}^{(2)}$, $h^{(1)}$, and $h^{(2)}$, and set the other parameters to $d_y^{(1)}=d_y^{(2)}=9$ mm as before. In order not to force any \textit{a priori} asymmetry in the groove configuration, we fix their centers to $(a^{(1)}_{x},a^{(1)}_{y})=(0.25L_{x},0.5L_{y}),(a^{(2)}_{x},a^{(2)}_{y})=(0.75L_{x},0.5L_{y})$. In view of the nonlinear nature of the constraints, it is clear that these settings are not unique, and other choices may lead to valid solutions as well.


The next step, thus, is to 
enforce the perfect anomalous reflection condition for the TM-polarized fields as before; this time, however, we need to set the aforementioned four DOFs such that  
$\eta^{m}_{1,0}\to 1$ [Eq. \eqref{eq:eta_condition}].
To resolve this nonlinear equation more efficiently, 
we use the library function \texttt{lsqnonlin} in MATLAB \cite{rabinovich2019arbitrary}, facilitating rapid assessment of these four geometrical parameters. 
For the particular case considered herein, one such valid solution prescribed the width and depth of the first groove as $d_x^{(1)}=7.92$ mm and $h^{(1)}=10.92$ mm, respectively, and indicated that the second groove should be wider and deeper, with $d_x^{(2)}=11.85$ mm and $h^{(2)}=19.94$ mm.

\begin{figure}[t]
\centering
\includegraphics[width=2.5in]{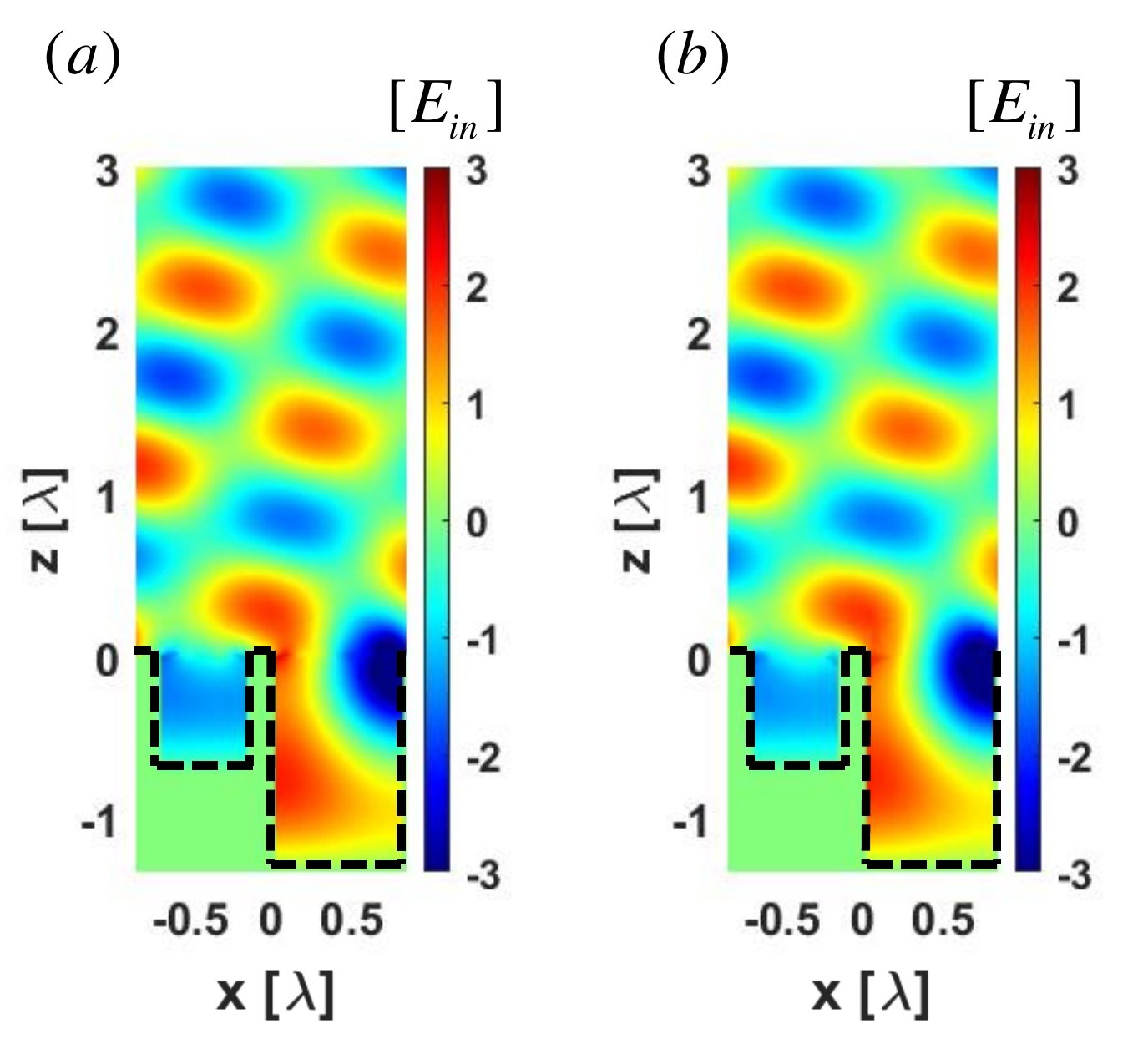}
\caption{Electric field distribution $\Re\left\{E_x(x,y=0,z)\right\}$ corresponding to the single-polarized anomalous reflector of Fig. \ref{Fig:Config_10to50}, when illuminated from $\theta_{\mathrm{in}}=10^{\circ}$ by a TM-polarized plane wave at $f=20$ GHz (single period is shown). 
The analytical prediction (a) following Eq. \eqref{eq:Matrix_equation} with Eqs. \eqref{eq:E_H_inc}-\eqref{eq:TM_TE_modes_holes}, is compared with full-wave simulation results (b). The black dashed lines mark the boundaries of the metallic construct. }
\label{Fig:Field_10to50}
\end{figure}

These values were used to define and simulate the prototype MG structure in CST Microwave Studio.
A comparison between the fields on the $\widehat{xz}$ plane as calculated in this full-wave simulation and the ones derived from the analytical model is shown in Fig. \ref{Fig:Field_10to50}, revealing, again, an excellent agreement between the two.
Simulated results predict extremely low ($0.3\%$) power coupling to the spurious $(-1,0)$ and $(0,0)$ FB harmonics, with $98\%$ of the incident power redirected towards the prescribed anomalous reflection mode at $\theta_\mathrm{out}=50.7^\circ$, and 
approximately $2\%$ absorption in the textured aluminium slab. 

With these encouraging results, we proceeded with fabrication of the prototype [Fig. \ref{Fig:Config_10to50} (b)]. As discussed in Section \ref{subsec:singl_pol_one_groove}, manufacturing constraints again yielded grooves with curved corners, with radius of curvature of $1.5$ mm (the deeper grooves and their close proximity dictated stricter constraints herein). Adapting the simulated structure accordingly, it was found that the anomalous reflection efficiency has somewhat deteriorated due to this deviation from the model, though still reaching a high value of $\eta^m_{\mathrm{tot}}=91.2\%$. 

\begin{figure}[t]
\centering
\includegraphics[width=2.5in]{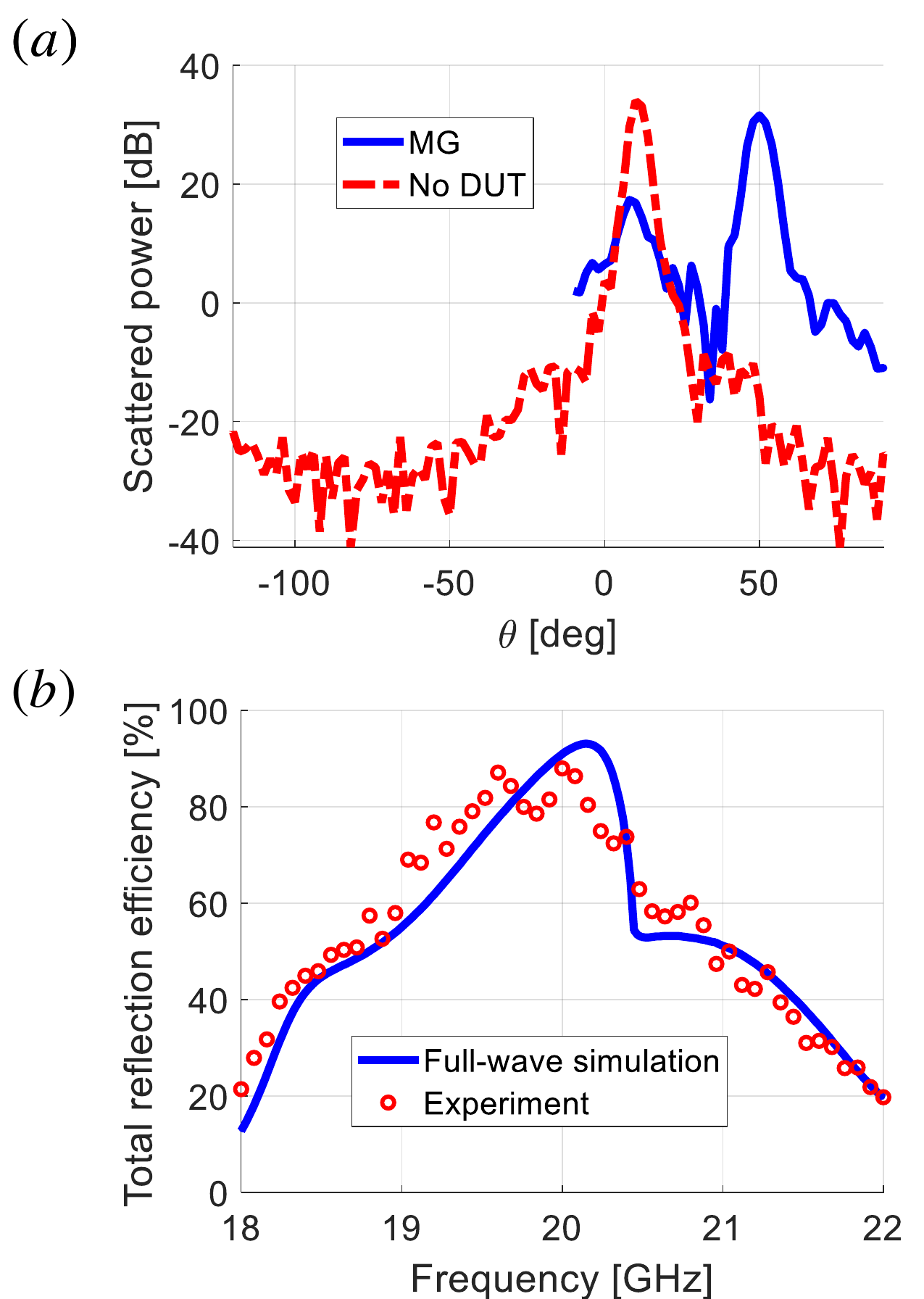}
\caption{Experimental characterization of the single-polarized anomalous reflection MG of Fig. \ref{Fig:Config_10to50}. (a) Received power ($f=20$ GHz) as a function of the observation angle $\theta$ (scattering pattern) as obtained when the MG was excited (solid blue) is compared to the reference pattern recorded in the absence of the DUT (dash-dotted red).
(b) Total anomalous reflection efficiency $\eta^m_\mathrm{tot}$ of the MG prototype towards the $(n_x,n_y)=(1,0)$ FB mode as a function of frequency. Experimental results (red circles) are compared with the ones obtained via full-wave simulation (solid blue).}
\label{Fig:EXP_10to_50_TM}
\end{figure}

The fabricated MG was subsequently tested in an anechoic chamber in the same manner described in Section \ref{subsec:singl_pol_one_groove} (Fig. \ref{Fig:EXP_Setup}). The measured scattering pattern (at $f=20$ GHz) associated with the MG is presented in Fig. \ref{Fig:EXP_10to_50_TM}(a) (solid blue line), along with the reference measurement of the Gaussian beam illumination (dash-dotted red), taken in the absence of the DUT.
It can be seen that at $\theta=50^{\circ}$, the MG reflection reaches its maximum (very close to the designated angle $\theta_{\mathrm{out}}=50.7^{\circ}$), while the specular reflection at $\theta=10^{\circ}$ is suppressed by more than $15$ dB relative to the reference. 
The anomalous reflection efficiency evaluated via Eq. \eqref{eq:efficiency_EXP} and presented in Fig. \ref{Fig:EXP_10to_50_TM}(b) as a function of frequency highlights, again, a very good agreement between full-wave simulations (solid blue) and measurements (red circles), with 
the trend of the experimental graph following relatively tightly after the numerically assessed curve. The total anomalous reflection efficiency measured at the designated operating frequency $f=20$ GHz is $\eta^m_\mathrm{tot}=88\%$, a mere $3.5\%$ relative deviation from the $91.2\%$ predicted in full-wave simulations. These results validate experimentally the ability of the presented synthesis procedure to design versatile efficient metal-based MGs with multiple grooves per period to control multiple diffraction orders. 

\subsection{Dual-polarized anomalous reflection}
\label{subsec:dual_pol}
\begin{figure}[t]
\centering
\includegraphics[width=3.2in]{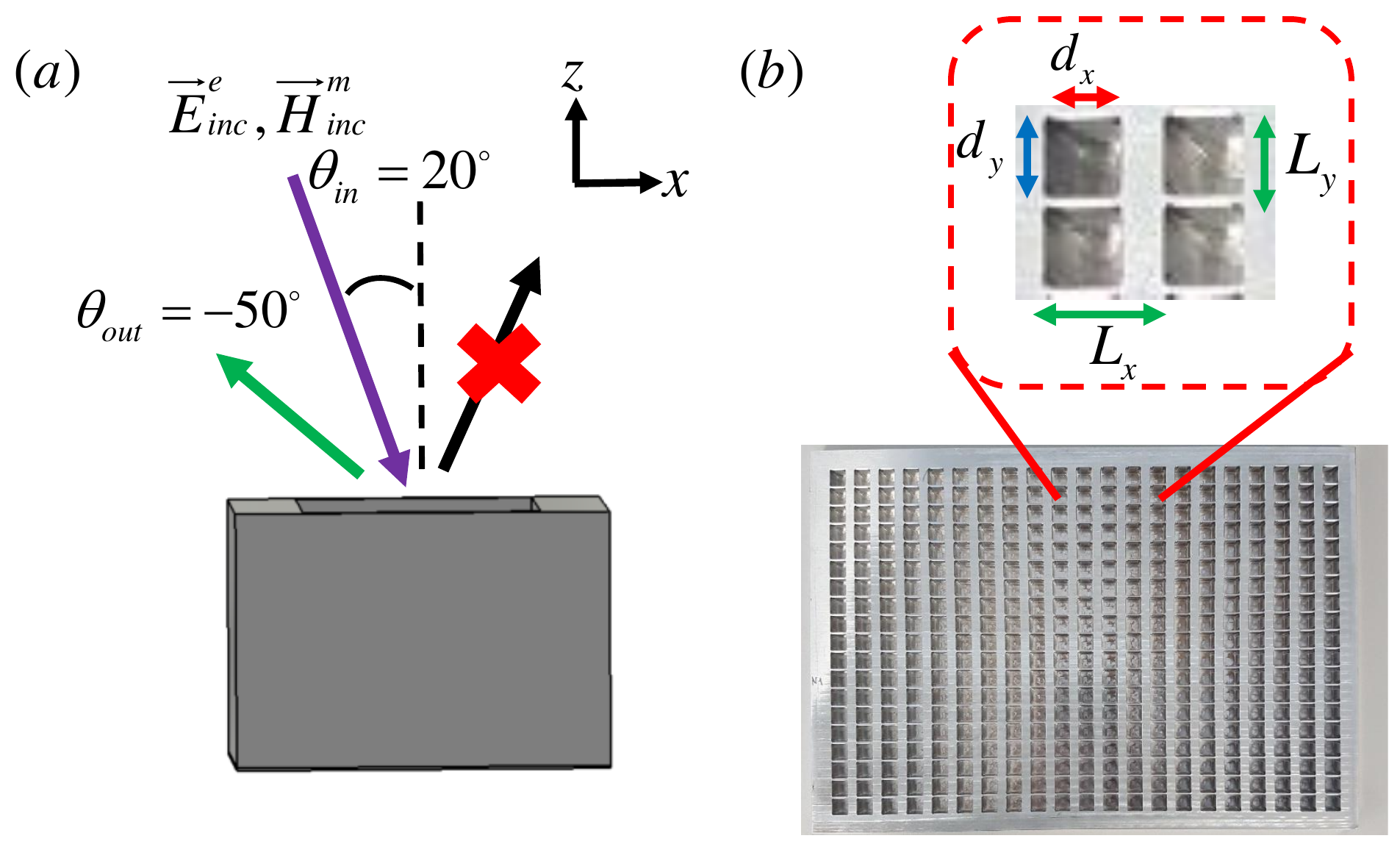}
\caption{MG designed for dual-polarized anomalous reflection from $\theta_{\mathrm{in}}=20^{\circ}$ towards $\theta_{\mathrm{out}}=-50^{\circ}$ at $f=20$ GHz, featuring two propagating FB modes. (a) Physical configuration (one groove per period). 
(b) Manufactured prototype. Inset: close-up on 4 unit cells; due to CNC fabrication limitations, $1$ mm curvature was introduced to the groove corners. }
\label{Fig:Config_20tominus50}
\end{figure}

In the last case study we consider herein, we wish to demonstrate the feasibility of the proposed methodology to design a dual-polarized all-metallic MG, implementing efficient polarization-insensitive anomalous reflection. Specifically, we designate the MG to deflect simultaneously both TM- and TE- polarized plane waves incoming from $\theta_{\mathrm{in}}=20^{\circ}$ towards $\theta_{\mathrm{out}}=-50^{\circ}$ at $f=20$ GHz, leading to periodicity of $L_{x}=\lambda/|\sin\theta_{\mathrm{in}}-\sin\theta_{\mathrm{out}}|=13.54$ mm $=0.9\lambda$ [Fig. \ref{Fig:Config_20tominus50}(a)]. In this scenario, as in Section \ref{subsec:singl_pol_one_groove}, only two FB modes are propagating, corresponding to the specular $(n_{x},n_{y})=(0,0)$ and anomalous $(-1,0)$-order reflection. Thus, to obtain perfect anomalous reflection for both polarized excitations, we need to suppress specular reflection for both TE- and TM-polarized incident fields, leading to two constraints as per Eq. \eqref{eq:eta_condition}: 
$\eta^{m}_{-1,0}\to 1$ and $\eta^{e}_{-1,0}\to 1$.

\begin{figure*}[t]
\centering
\includegraphics[width=4.5in]{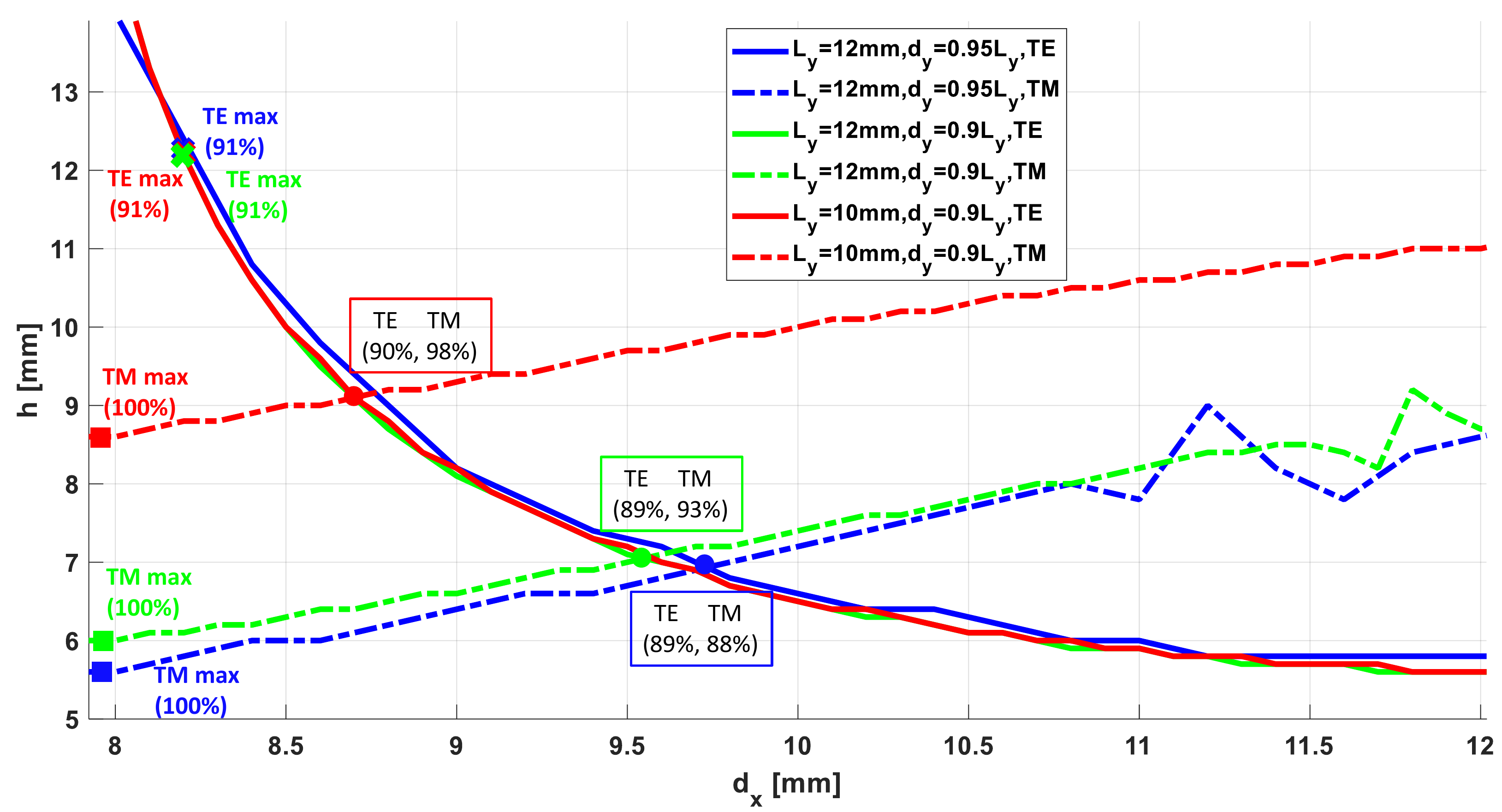}
\caption{Maximum-efficiency curves for the dual-polarized MG. For each of the considered $(L_y,d_y)$ combinations, namely, $(L_y=10\mathrm{mm},d_y=0.9L_y)$ (red), $(L_y=12\mathrm{mm},d_y=0.9L_y)$ (green), and $(L_y=12\mathrm{mm},d_y=0.95L_y)$ (blue), solid and dash-dotted lines denote, respectively, the groove depth $h$ providing the best anomalous reflection efficiency for the TE and TM excitation scenarios, as a function of the groove width $d_x$.
Intersection points between maximum-efficiency curves of the same color are marked by a circle, denoting a potential working point for the dual-polarized device. Square and X markers denote, respectively, optimal operating conditions for TM- and TE- single-polarized anomalous reflectors. Textboxes adjacent to the marked points present the expected anomalous reflection efficiencies.}
\label{Fig:eta_20to_minus50_TM_TE_Ly_dy}
\end{figure*}

For the single-polarization MG designed for the two-channel scenario of Section \ref{subsec:singl_pol_one_groove}, we utilized two DOFs, $d_x$ and $h$, corresponding to a MG with a single groove per period. However, since the dual-polarized scenario introduces an additional constraint, we choose to put into play the DOFs stemming from the configuration's variation along the $y$ direction, $L_y$ and $d_y$, which we have yet to harness, still considering one groove in the period. To assess the potential performance achievable with this extended set of DOFs, we consider three representative combinations of $(L_y,d_y)$ for each of the polarizations, and examine graphically the possibility to reach efficient dual-polarized anomalous reflection. More specifically, for each of these combinations, we sweep the value of $d_x$ and mark these depths $h$ that lead to the maximal anomalous reflection efficiency for either TE- or TM-  polarized excitations. Combining these $(d_x,h)$ points together for a given $(L_y,d_y)$ forms "maximum efficiency" curves for each of the polarizations, plotted in Fig. \ref{Fig:eta_20to_minus50_TM_TE_Ly_dy}. Therein, solid lines and dash-dotted lines denote, respectively, maximum-efficiency curves for the TE (highest $\eta^e_{-1,0}$) and TM (highest $\eta^m_{-1,0}$) scenarios.

As the MG should perform the anomalous reflection with high efficiency for both polarization cases simultaneously, we consider intersection points of these graphs as valid solutions for dual-polarized operation. For each such intersection point [denoted by a circle, filled with color according to the associated $(L_y,d_y)$ combination], we indicated in a rectangular textbox the respective anomalous reflection efficiencies obtained from the analytical model for the corresponding MG configuration. In addition, we marked on each of the curves the working point for which the best single-polarized anomalous reflection efficiency was achieved for the particular polarization relevant to this curve: X markers for TE-related curves (solid lines), and squares for the TM scenarios (dash-dotted); adjacent numerical values stand for the maximal efficiency recorded ($\eta^e_{-1,0}$ or $\eta^m_{-1,0}$, respectively).

As can be seen, due to the need to obtain simultaneously high efficiencies for both polarizations, the dual-polarized working points may yield anomalous reflection efficiencies which are smaller than the optimal values achievable for each polarization independently. However, since we can tune the geometrical DOFs of the groove in both lateral dimensions, an operating point with minimal reduction in performance can be found, e.g. the one marked by a red circle, corresponding to $L_y=10$ mm, $d_y=9$ mm, $d_x=8.6$ mm, and $h=9.2$ mm. This highlights the importance of the 2D rectangular groove configuration utilized herein for obtaining efficient dual-polarized operation with one groove per period. Without the ability to modify the MG geometry along both the $x$ and $y$ axes, greater compromise with respect to the single-polarized performance may be inevitable.

It is also interesting to note that the achievable TE-polarized performance of the MG as reflected in Fig. \ref{Fig:eta_20to_minus50_TM_TE_Ly_dy} is hardly affected by the variation of $d_y$ and $L_y$. In particular, the TE anomalous reflection efficiency recorded along the maximum-efficiency curves remains around $\sim 90\%$. This "stability" implies that the TE-polarized incident fields predominantly couple to the $y$-invariant mode in the grooves, which is always above cutoff (for the relevant range of $d_x$) for TE ($\vec{E}^e \propto \hat{y}$) guided modes \footnote{In contrast to TM-polarized guided modes, which are always affected by $d_y$ via the propagation constant.}.

Indeed, it seems from Fig. \ref{Fig:eta_20to_minus50_TM_TE_Ly_dy} that the TE anomalous reflection is limited in efficiency: in contrast to the anoamlous reflection for TM-polarized incident fields, which reaches near-unitary efficiencies for optimal groove dimensions, no working conditions within the large parameter space we explored led to TE-polarized anomalous reflection with more than $91\%$. 
To further examine this phenomenon and its dependency on the scattering scenario, we fixed the angle of incidence to $\theta_{\mathrm{in}}=30^{\circ}$ and considered anomalous reflection towards a variety of angles $\theta_\mathrm{out}$ between $-30^{\circ}$ and $-90^{\circ}$ (all these cases feature only two radiation channels). For each deflection angle $\theta_\mathrm{out}$, we assessed based on the analytical model the best possible anomalous reflection efficiency obtainable for TE-polarized incident fields, considering a wide range of $(d_x, h)$ combinations to this end \footnote{Considering the relative invariance of the TE anomalous reflection peak efficiencies with respect to the geometrical parameters associated with the $y$-axis dimensions of the MG, we kept $L_y$ and $d_y$ constant during this parametric study.}.

\begin{figure}[t]
\centering
\includegraphics[width=2.5in]{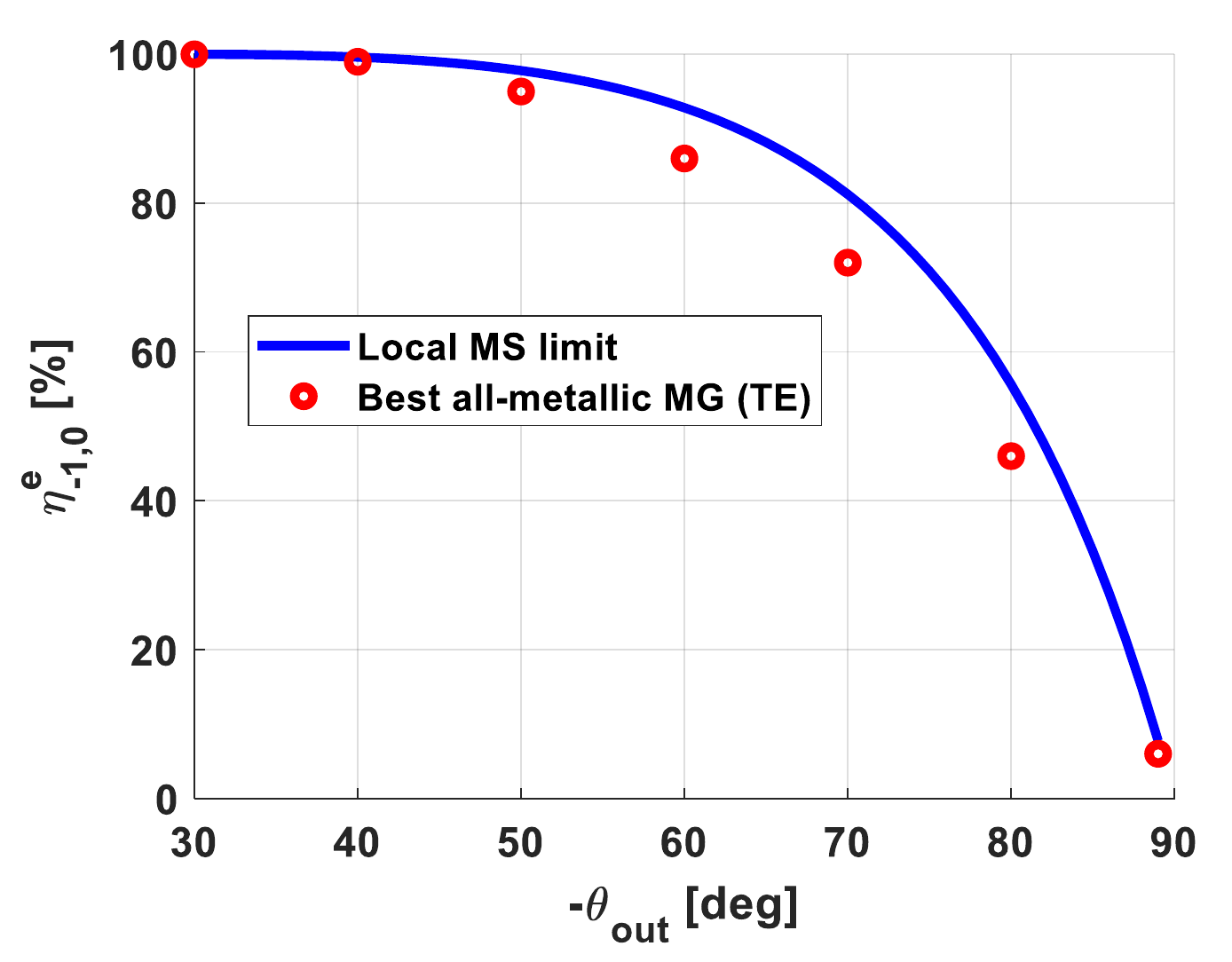}
\caption{Comparison between the best TE anomalous reflection efficiency achievable for a single-groove all-metallic MG (red circles) and the theoretical limit for local impenetrable MSs \cite{asadchy2016perfect} (solid blue) as a function of $\theta_\mathrm{{out}}$ ($\theta_\mathrm{{in}}=30^{\circ}$).}
\label{Fig:Max_eff_TE}
\end{figure}

The resultant maximum-efficiency points are marked by red circles in Fig. \ref{Fig:Max_eff_TE}. Remarkably, the emerging angular dependency closely follows the theoretical limit derived in \cite{asadchy2016perfect} for \emph{local} impenetrable anomalous reflection MSs (blue solid line). For such devices, specularly reflected fields with a reflection magnitude of $\Gamma=[\cos(\theta_\mathrm{out})-\cos(\theta_\mathrm{in})]/[\cos(\theta_\mathrm{out})+\cos(\theta_\mathrm{in})]$ are required in order to satisfy local power conservation and local impedance equalization, facilitating passive lossless design without macroscopic non-locality \cite{selvanayagam2013discontinuous, epstein2016huygens}. In order to overcome this limitation of $\eta^{\mathrm{max}}_{-1,0}=1-\left|\Gamma\right|^2$, incurring spurious reflections that become more severe as the wave-impedance mismatch between the incident and reflected waves increases, it is required to introduce some means by which power could be transferred in the transverse direction along the surface, such as auxiliary surface waves \cite{epstein2016synthesis, diaz2017generalized, asadchy2017eliminating}. Thus, the similarity revealed in Fig. \ref{Fig:Max_eff_TE} implies that for TE-polarized incident fields, wide-angle anomalous reflection with all-metallic MGs is limited due to challenges in excitation of such surface waves \footnote{In previously investigated MGs, it is this ability to form macroscopic non-locality via excitation of transversally propagating evanescent modes which allowed demonstration of ultra-high anomalous reflection efficiencies \cite{ra2017meta, epstein2017unveiling, wong2018perfect}.}. 

\begin{figure}[t]
\centering
\includegraphics[width=2.5in]{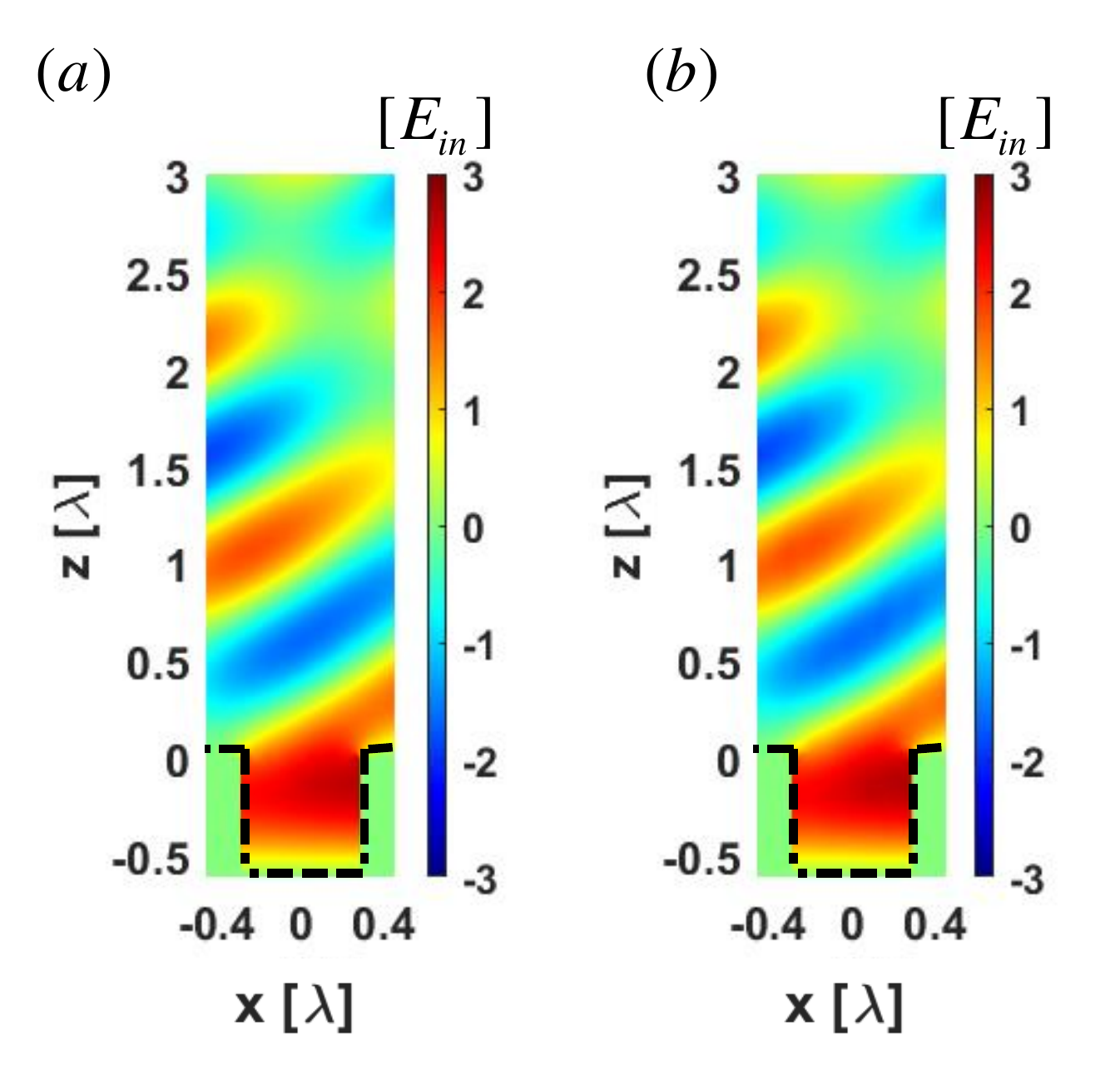}
\caption{Electric field distribution $\Re\left\{E_x(x,y=0,z)\right\}$ corresponding to the dual-polarized anomalous reflector of Fig. \ref{Fig:Config_20tominus50}, when illuminated from $\theta_{\mathrm{in}}=20^{\circ}$ by a TM-polarized plane wave at $f=20$ GHz (single period is shown). 
The analytical prediction (a) following Eq. \eqref{eq:Matrix_equation} with Eqs. \eqref{eq:E_H_inc}-\eqref{eq:TM_TE_modes_holes}, is compared with full-wave simulation results (b). The black dashed lines mark the boundaries of the metallic construct.}
\label{Fig:Field_TM_20tominus50}
\end{figure}

Indeed, it seems to be more difficult to generate and guide surface waves for this polarization on certain all-metallic structures \cite{collin1991field}. 
Overall, this investigation indicates that while dual-polarized retroreflection with unitary efficiency \cite{Rabinovich2020retro} can be readily achieved with the proposed configuration, retaining this optimal performance for TE-polarized fields becomes more challenging as the wave-impedance mismatch between incident and reflected waves grows larger. Nonetheless, for moderate-angle anomalous reflection such as the one demonstrated in this subsection ($\theta_\mathrm{in}=20^\circ$, $\theta_\mathrm{out}=-50^\circ$), very high efficiency ($>90\%$) can still be achieved simultaneously for both TE- and TM-polarized excitations, due to the ability to tune the $y$-axis dimensions of the MG (Fig. \ref{Fig:eta_20to_minus50_TM_TE_Ly_dy}). As indicated in \cite{collin1991field}, 
the use of high-index dielectrics (instead of vacuum) inside the grooves may mitigate this issue to an extent.

\begin{figure}[t]
\centering
\includegraphics[width=2.3in]{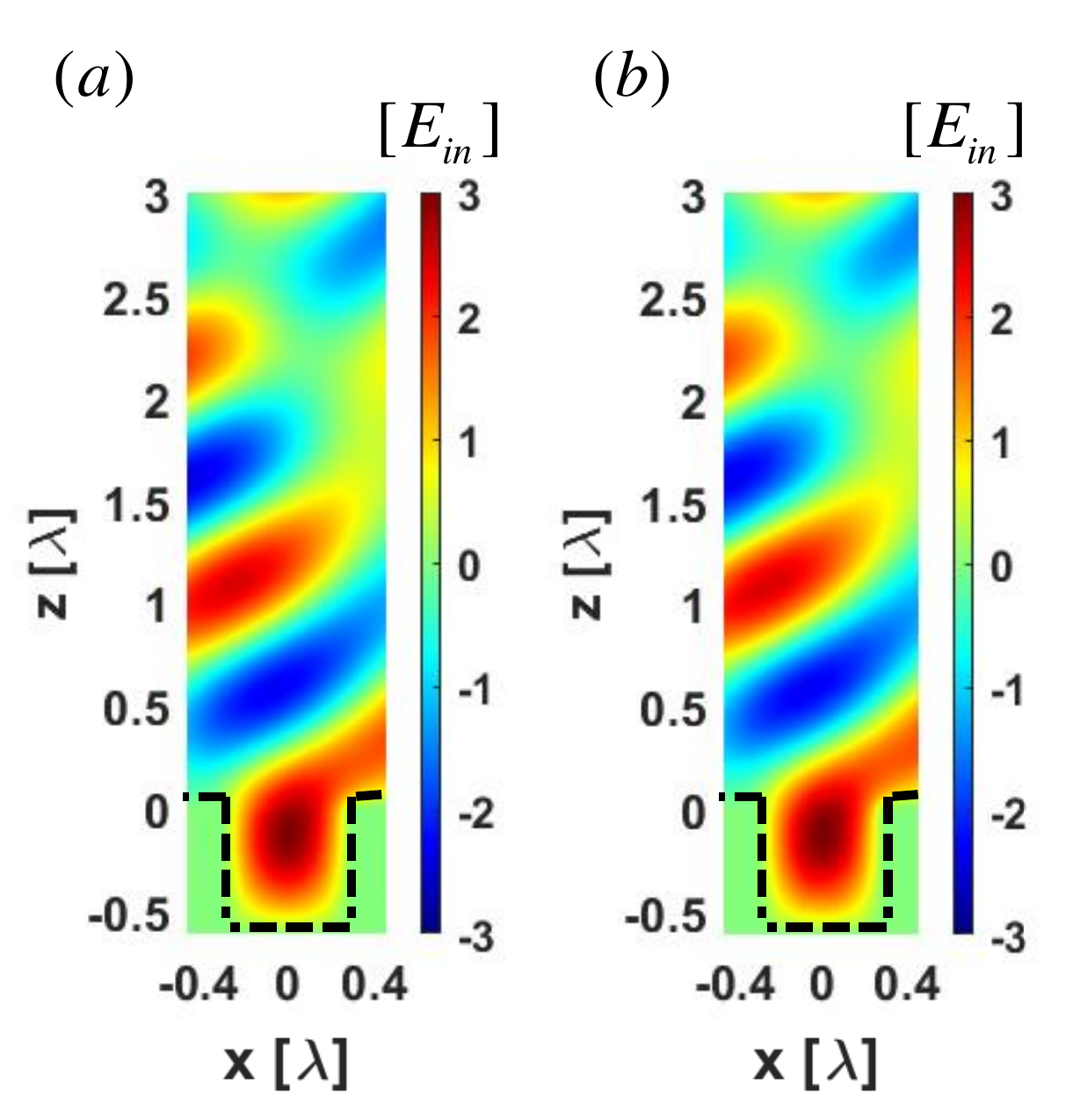}
\caption{Electric field distribution $\Re\left\{E_x(x,y=0,z)\right\}$ corresponding to the dual-polarized anomalous reflector of Fig. \ref{Fig:Config_20tominus50}, when illuminated from $\theta_{\mathrm{in}}=20^{\circ}$ by a TE-polarized plane wave at $f=20$ GHz (single period is shown). 
The analytical prediction (a) following Eq. \eqref{eq:Matrix_equation} with Eqs. \eqref{eq:E_H_inc}-\eqref{eq:TM_TE_modes_holes}, is compared with full-wave simulation results (b). The black dashed lines mark the boundaries of the metallic construct.}
\label{Fig:Field_TE_20tominus50}
\end{figure}

Returning to the problem at hand, we use the working point marked by a red circle in Fig. \ref{Fig:eta_20to_minus50_TM_TE_Ly_dy} for the dual-polarized device, and define the resultant MG in CST Microwave Studio for verification. The comparison between the electric fields as predicted by the analytical model and as recorded in full-wave simulations is presented in  Figs. \ref{Fig:Field_TM_20tominus50} and \ref{Fig:Field_TE_20tominus50} for the TM- and TE- polarized incident plane waves, respectively.
Once more, an excellent agreement is observed in both cases. Full-wave simulations indicate that $98\%$ of the TM-polarized and $90\%$ of the TE-polarized incident plane-wave power are coupled to the anomalous reflection mode, while the rest is specularly reflected (consistent with the discussion related to Fig. \ref{Fig:Max_eff_TE}). Even though the TE-polarized anomalous reflection efficiency is somewhat lower due to the limitations discussed in the previous paragraph, the resultant MG design still serves as a good example for simultaneous high-efficiency beam deflection control at two polarizations.

As in the previous case studies, the prescribed MG specifications were used to fabricate a prototype for experimental validation [Fig. \ref{Fig:Config_20tominus50} (b)]. The machining process constraints introduced, as in Section \ref{subsec:singl_pol_one_groove}, a radius of curvature of $1$ mm to the rectangular groove corners. Rerunning the full-wave simulations with the deformed grooves pointed out that the peak efficiency did not decrease, but a minor frequency shift was caused to the TE anomalous reflection curves, yielding the best performance at $f=20.3$ GHz.

\begin{figure}[t]
\centering
\includegraphics[width=2.5in]{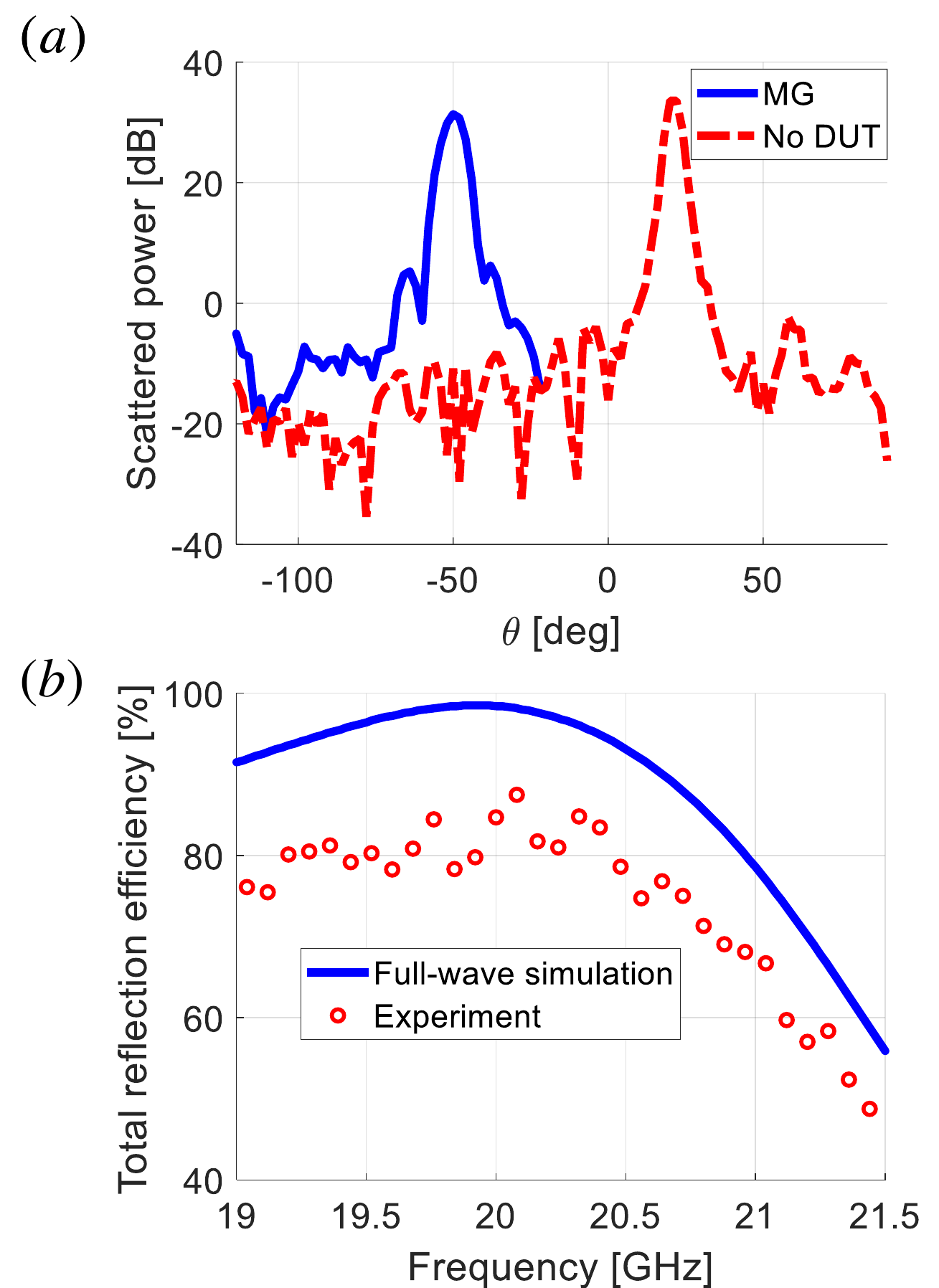}
\caption{Experimental characterization of the dual-polarized anomalous reflection MG of Fig. \ref{Fig:Config_20tominus50}. (a) Received power ($f=20$ GHz) as a function of the observation angle $\theta$ as obtained when the MG was excited by the TM-polarized Gaussian beam (solid blue) is compared to the reference pattern recorded in the absence of the DUT (dash-dotted red).
(b) Total anomalous reflection efficiency $\eta^m_\mathrm{tot}$ of the MG prototype towards the $(n_x,n_y)=(-1,0)$ FB mode as a function of frequency. Experimental results (red circles) are compared with the ones obtained via full-wave simulation (solid blue).}
\label{Fig:EXP_20to_minus50_TM}
\end{figure}

\begin{figure}[t]
\centering
\includegraphics[width=2.5in]{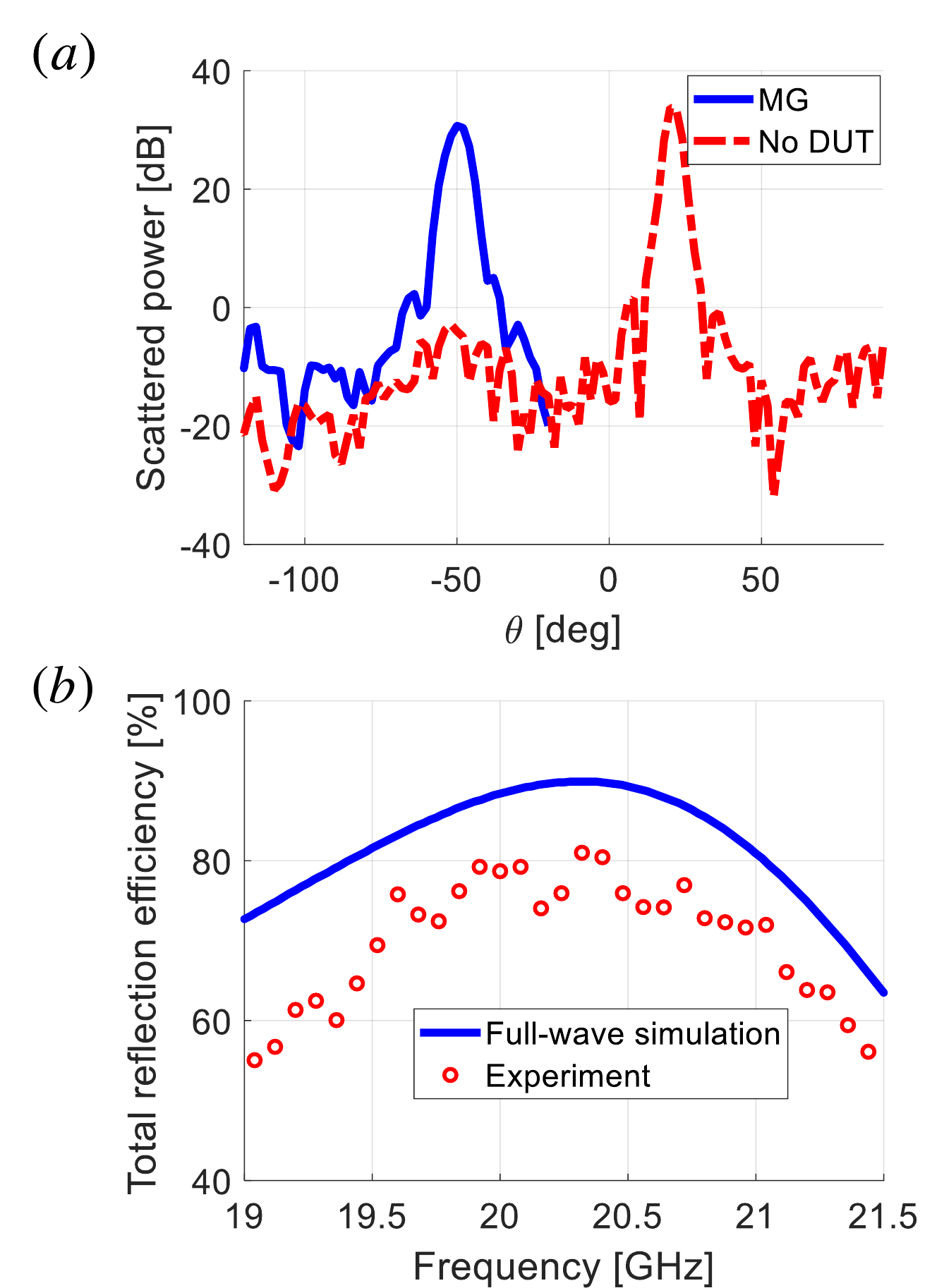}
\caption{Experimental characterization of the dual-polarized anomalous reflection MG of Fig. \ref{Fig:Config_20tominus50}. (a) Received power ($f=20$ GHz) as a function of the observation angle $\theta$ as obtained when the MG was excited by the TE-polarized Gaussian beam (solid blue) is compared to the reference pattern recorded in the absence of the DUT (dash-dotted red).
(b) Total anomalous reflection efficiency $\eta^e_\mathrm{tot}$ of the MG prototype towards the $(n_x,n_y)=(-1,0)$ FB mode as a function of frequency. Experimental results (red circles) are compared with the ones obtained via full-wave simulation (solid blue).}
\label{Fig:EXP_20to_minus50_TE}
\end{figure}

The fabricated prototype was tested again using our cylindrical near-field measurement system. This time, however, excitation and scattering at both polarizations were considered, by suitably rotating the Gaussian beam antenna and probe by $90^\circ$ around their axes. The measured scattering patterns corresponding to TM- and TE- polarized incident fields are presented in Fig. \ref{Fig:EXP_20to_minus50_TM}(a) and Fig. \ref{Fig:EXP_20to_minus50_TE}(a), respectively, where the beam deflection towards $\theta_\mathrm{out}=-50^\circ$ can be clearly seen.

The anomalous reflection efficiencies for the two polarizations as evaluated from the experimental data are depicted in Fig. \ref{Fig:EXP_20to_minus50_TM}(b) and Fig. \ref{Fig:EXP_20to_minus50_TE}(b), along with the simulation predictions. While the measured frequency dependency follows quite closely the simulated one, we identify a rather constant drop in the efficiency values of about $10\%$. In particular, the measured efficiency for the TM-polarized and TE-polarized anomalous reflection peak at $\eta^{m}_{\mathrm{tot}}=88\%$ and $\eta^{e}_{\mathrm{tot}}=81\%$, instead of the theoretical predictions of $\eta^{m}_{\mathrm{tot}}=98\%$ and $\eta^{e}_{\mathrm{tot}}=90\%$ (Figs. \ref{Fig:Field_TM_20tominus50} and \ref{Fig:Field_TE_20tominus50}). 
This discrepancy is attributed to the blockage issue mentioned in Section \ref{subsec:singl_pol_one_groove}, which becomes more pronounced in the current scenario, in which the incident and anoamlous reflection angles ($\theta_{\mathrm{in}}$ and $\theta_{\mathrm{out}}$) are closer to one another. Consequently, a portion of the power reflected towards $\theta_{\mathrm{out}}$
actually diffracts off the edges of the Gaussian beam antenna and is not recorded in the near-field probe (Fig. \ref{Fig:EXP_Setup}). In fact, this problem has been more acute in the original measurement setup, where the Gaussian beam antenna is positioned quite close to the sample ($196$ mm $\approx$ 13 $\lambda$). To reduce the blockage issue, we have increased the distance between the excitation antenna and the MG to $450$ mm $\approx 30 \lambda$, which meets the limits of our apparatus; the results presented in Figs. \ref{Fig:EXP_20to_minus50_TM} and \ref{Fig:EXP_20to_minus50_TE} were recorded using this improved setup. 
%
Overall, the combined verification, via full-wave simulations and laboratory measurements, demonstrate the accuracy of the presented analytical model, and its effectiveness for the realization of highly-efficient MG devices. 

\section{Conclusion}
\label{sec:conclusion}

\begin{table*}[t]
\centering
\begin{threeparttable}[b]
\renewcommand{\arraystretch}{1.3}
\caption{The modal coefficients ($\bm{n}$th FB mode for $\alpha_{\bm{n}}^{(p)},\beta_{\bm{n}}^{(p)}$ and $\bm{m}$th rectangular waveguide mode in the $i$th groove for $\Gamma_{\bm{m}}^{(p,i)},\Delta_{\bm{m}}^{(p,i)}$) and the excitation source coefficients ($S_{\bm{n}}^{(p)}$ for the $\bm{n}$th FB mode and $S_{\bm{m}}^{(p)}$ for the $\bm{m}$th rectangular waveguide mode) presented in equations \eqref{eq:Final_equation_Ex}-\eqref{eq:Final_equation_Ey} and in \eqref{eq:Final_equation_Hx}-\eqref{eq:Final_equation_Hy}, evaluated using the boundary conditions at $z=0$.}
\label{tab:impedance_matrix_coefficients}
\centering

\begin{tabular}{c||c|c|c|c|c|c}
\hline \hline

& $\alpha_{\bm{n}}^{(p)}$ & $\beta_{\bm{n}}^{(p)}$ & $\Gamma_{\bm{n}}^{(p,i)}$ & $\Delta_{\bm{n}}^{(p,i)}$ & $S_{\bm{n}}^{(p)}$ or $S_{\bm{m}}^{(p,i)}$ 
 \\ 
\hline \hline \\[-1.3em]
	\begin{tabular}{c} $p=1$ \\ \footnotesize{[Eq. \eqref{eq:Final_equation_Ex}]}  \end{tabular}
	 & $\frac{-\eta k_{x}k_{z}}{k_{x}^{2}+k_{y}^{2}}$ 
	 & $\frac{-\eta k k_{y}}{k_{x}^{2}+k_{y}^{2}}$ 
	 & $\frac{j\eta_{d}^{(i)}k_{z,grv}^{(i)} k_{x,grv}^{(i)} p_{m_x,m_y} \sinh(jk_{z,grv}^{(i)}h^{(i)})}{[k_{x,grv}^{(i)}]^{2}+[k_{y,grv}^{(i)}]^{2}}$ 
	 & $\frac{j\eta k k_{y,grv}^{(i)}q_{m_x,m_y}\sinh(jk_{z,grv}^{(i)}h^{(i)})}{[k_{x,grv}^{(i)}]^{2}+[k_{y,grv}^{(i)}]^{2}}$  
	 & $s_{\bm{n}}^{\{e,m\},(1)}\delta_{\bm{n},\bm{0}}$  
	  \\	\hline	
	 \begin{tabular}{c} $p=2$ \\ \footnotesize{[Eq. \eqref{eq:Final_equation_Ey}]} \end{tabular}
	 & $\frac{-\eta k_{y}k_{z}}{k_{x}^{2}+k_{y}^{2}}$ 
	 & $\frac{\eta k k_{x}}{k_{x}^{2}+k_{y}^{2}}$ 
	 & $\frac{j\eta_{d}^{(i)}k_{z,grv}^{(i)} k_{y,grv}^{(i)} p_{m_x,m_y} \sinh(jk_{z,grv}^{(i)}h^{(i)})}{[k_{x,grv}^{(i)}]^{2}+[k_{y,grv}^{(i)}]^{2}}$ 
	 & $-\frac{j\eta k k_{x,grv}^{(i)}q_{m_x,m_y}\sinh(jk_{z,grv}^{(i)}h^{(i)})}{[k_{x,grv}^{(i)}]^{2}+[k_{y,grv}^{(i)}]^{2}}$  
	 & $s_{\bm{n}}^{\{e,m\},(2)}\delta_{\bm{n},\bm{0}}$  
	  \\	\hline 
	 \begin{tabular}{c} $p=3$ \\ \footnotesize{[Eq. \eqref{eq:Final_equation_Hx}]} \end{tabular}
	 & $\frac{\eta k k_{y}}{k_{x}^{2}+k_{y}^{2}}$ 
	 & $\frac{-\eta k_{x}k_{z}}{k_{x}^{2}+k_{y}^{2}}$ 
	 & $\frac{j\cdot \eta_{d}^{(i)} \varepsilon_{d}^{(i)} k\cdot k_{y,grv}^{(i)}p_{m_x,m_y}\cosh(jk_{z,grv}^{(i)}h^{(i)})}{[k_{x,grv}^{(i)}]^{2}+[k_{y,grv}^{(i)}]^{2}}$ 
	 & $-\frac{j\eta\cdot{k_{x,grv}^{(i)}k_{z,grv}^{(i)}}q_{m_x,m_y}\cosh(jk_{z,grv}^{(i)}h^{(i)})}{[k_{x,grv}^{(i)}]^{2}+[k_{y,grv}^{(i)}]^{2}}$ 
	 & $\sum\limits_{n=-\bm{N}/2}^{\bm{N}/2}\!\!\!\!\!\!s_{\bm{n}}^{\{e,m\},(3)}\delta_{\bm{n},\bm{0}}\chi^{(i)*}_{\bm{n,m}}$ 
	  \\	\hline 	  
	 \begin{tabular}{c} $p=4$ \\ \footnotesize{[Eq. \eqref{eq:Final_equation_Hy}]} \end{tabular}
	 & $\frac{-\eta k k_{x}}{k_{x}^{2}+k_{y}^{2}}$
	 & $\frac{-\eta k_{y}k_{z}}{k_{x}^{2}+k_{y}^{2}}$ 
	 & $-\frac{j\cdot \eta_{d}^{(i)} \varepsilon_{d}^{(i)} k\cdot k_{x,grv}^{(i)}p_{m_x,m_y}\cosh(jk_{z,grv}^{(i)}h^{(i)})}{[k_{x,grv}^{(i)}]^{2}+[k_{y,grv}^{(i)}]^{2}}$ 
	 & $-\frac{j\eta\cdot{k_{y,grv}^{(i)}k_{z,grv}^{(i)}}q_{m_x,m_y}\cosh(jk_{z,grv}^{(i)}h^{(i)})}{[k_{x,grv}^{(i)}]^{2}+[k_{y,grv}^{(i)}]^{2}}$ 
	 & $\sum\limits_{n=-\bm{N}/2}^{\bm{N}/2}\!\!\!\!\!\!s_{\bm{n}}^{\{e,m\},(4)}\delta_{\bm{n},\bm{0}}\psi^{(i)*}_{\bm{n,m}}$ 
	  \\		
\hline \hline
\end{tabular} 

\end{threeparttable}
\end{table*}

To conclude, we have presented a mode-matching based analytical formalism and design scheme for the synthesis of all-metallic MGs for anomalous reflection. The proposed MG consists of a periodic arrangement of rectangular grooves in a metallic medium, possibly with multiple grooves per period, judiciously shaped and distributed as to implement the 
desired functionality. 
Following the typical MG synthesis approach, the detailed fabrication-ready device layout is obtained by applying suitable constraints on the semianalytically derived scattering coefficients, retrieving directly the suitable geometrical DOFs of the MG configuration. 

The efficacy and versatility of this approach have been demonstrated theoretically and experimentally using three prototypical case studies, realizing highly-efficient wide-angle anomalous reflection for TM-polarized waves with two and three radiation channels (with one and two grooves per period, respectively), as well as dual-polarized beam deflection, performing the prescribed functionality simultaneously for TE- and TM-polarized excitations. 
In particular, we highlighted the importance of the finite area of the 2D rectangular grooves (as opposed to 1D elongated corrugations) as a means to provide additional DOFs for facilitating dual-polarized operation; and the limitations encountered when targeting extreme anomalous reflection for TE-polarized fields, related to challenges in excitation and guidance of TE-polarized surface waves along metallic structures. 
These results provide insight into the possibility of all-metallic MGs to realize polarization sensitive and insensitive functionalities, as well as lay out practical engineering tools for designing efficient anomalous reflection devices for applications where dielectric-free structures are preferable.

\section*{Acknowledgement}
This research was supported by the Israel Science Foundation (Grant No. 1540/18). The authors gratefully acknowledge helpful discussions with Prof. Levi Sch\"{a}chter. They would also like to thank Yuri Komarovsky of the Communication Laboratory and Kalman Maler from the Mechanical Workshop in the Faculty for Electrical Engineering at the Technion for their assistance with regard to the experimental setup and device fabrication. Finally, they wish to thank the team of MVG/Orbit-FR in Israel for continuous technical support regarding acquisition and analysis with the near-ﬁeld measurement system.

\appendix*
\section{Boundary condition coefficients and overlap integrals}
\label{sec:Appendix}

This appendix provide explicit expressions for various parameters defined in Section \ref{sec:theory}, required for the stipulation and resolution of the boundary conditions related to the all-metallic MG configuration considered herein. In particular, the coefficients $\alpha_{\bm{n}}^{(p)},\beta_{\bm{n}}^{(p)},\Gamma_{\bm{m}}^{(p,i)},\Delta_{\bm{m}}^{(p,i)}, S^{(p)}_{\bm{n}},$ and $S^{(p,i)}_{\bm{m}}$ defined in Eqs. \eqref{eq:Boundary_conditions_Ex}-\eqref{eq:Final_equation_Hy}, needed for the evaluation of the matrix equation [Eq. \eqref{eq:Matrix_equation}] are given in Table \ref{tab:impedance_matrix_coefficients}, where $k_{x,\mathrm{grv}}^{(i)}=\pi m_x/d^{(i)}_{x}, k_{y,\mathrm{grv}}^{(i)}=\pi m_y/d^{(i)}_{y}$, and
\begin{equation}
\label{eq:p_q_condition}
\begin{aligned}
&p_{m_x,m_y}=\begin{cases}
0 &    \,\,\,\,\,\,m_x=0\vee m_y=0\\
    1 & \,\,\,\,\,\,     \text{otherwise}
\end{cases}\\
&q_{m_x,m_y}=\begin{cases}
0 &   \,\,\,\,\,\, m_x=0    \wedge  m_y=0\\
    1 &      \,\,\,\,\,\, \text{otherwise}
\end{cases}\\
\end{aligned}
\end{equation}
The polarization-dependent source terms are given by
\begin{equation}
\label{eq:source_coefficients}
\begin{array}{l l}
\!\!\!\!\!\!s_{\bm{n}}^{m,(1)}=\eta H^{m}_{0}\cos\theta_{\mathrm{in}} & s_{\bm{n}}^{e,(1)}=0 \\
\!\!\!\!\!\!s_{\bm{n}}^{m,(2)}=0 & s_{\bm{n}}^{e,(2)}=E_{0}^{e} \\
\!\!\!\!\!\!s_{\bm{n}}^{m,(3)}=0 & s_{\bm{n}}^{e,(3)}=\frac{k_{z}}{k}E_{0}^{e} \\
\!\!\!\!\!\!s_{\bm{n}}^{m,(4)}=-\frac{\eta H_{0}^{m}}{k}[k_z \cos\theta_{\mathrm{in}}+k_x \sin\theta_{\mathrm{in}}] & s_{\bm{n}}^{e,(4)}=0
\end{array}
%
\end{equation}

In addition, one can harness the integral identities
\begin{equation}
\label{eq:Analytical_exp}
\begin{aligned}
&\int_{v_{-}}^{v_{+}}e^{av}\sin(bv) dv = \left.\frac{e^{av}}{a^{2}+b^{2}}\cdot\big[a\sin(bv)-b\cos(bv)\big]\right|_{v_{-}}^{v_+} \\
&\int_{u_{-}}^{u_{+}}e^{cu}\cos(du) du = \left.\frac{e^{cu}}{c^{2}+d^{2}}\cdot\big[c\cos(du)+d\sin(du)\big]\right|_{u_{-}}^{u_+}
\end{aligned}
\end{equation}
to derive closed-form analytical expressions for the overlap integrals $\psi^{(i)}_{\bm{n},\bm{m}}$ and $\chi^{(i)}_{\bm{n},\bm{m}}$ of Eqs. \eqref{eq:psi} and \eqref{eq:chi}. Specifically, the expression for $\psi^{(i)}_{\bm{n},\bm{m}}$ is obtained from Eq. \eqref{eq:Analytical_exp} by substituting ${v=y},\, {v_-=y_-^{(i)}},\,{v_+=y_+^{(i)}},\,{a=jk_y},\,{b=\frac{\pi m_y}{d_y^{(i)}}(y-y_-^{(i)})}$, ${u=x},\, {u_-=x_-^{(i)}},\,{u_+=x_+^{(i)}},\,{c=jk_x},\,{d=\frac{\pi m_x}{d_x^{(i)}}(x-x_-^{(i)})}$, while $\chi^{(i)}_{\bm{n},\bm{m}}$ is obtained from Eq. \eqref{eq:Analytical_exp} by substituting ${v=x},\, {v_-=x_-^{(i)}},\,{v_+=x_+^{(i)}},\,{a=jk_x},\,{b=\frac{\pi m_x}{d_x^{(i)}}(x-x_-^{(i)})}$, ${u=y},\, {u_-=y_-^{(i)}},\,{u_+=y_+^{(i)}},\,{c=jk_y},\,{d=\frac{\pi m_y}{d_y^{(i)}}(y-y_-^{(i)})}$.
\end{document}